\documentclass[
preprint,
aps,prd,
nofootinbib,
superscriptaddress,
showpacs,
tightenlines,
amsmath,
amssymb
]{revtex4}
\usepackage{bm}
\usepackage[dvips]{color,graphicx}

\begin{document}
\title{Generalized Parton Distributions and Color Transparency Phenomena}

\author{S.~Liuti}
\email[]{sl4y@virginia.edu}

\author{S.~K.~Taneja}
\email[]{skt6c@virginia.edu}
\affiliation{University of Virginia, Charlottesville, Virginia 22904-4714, USA.}

\begin{abstract}
We study the structure of generalized parton distributions in 
impact parameter space with the aim of determining the size 
and role of small transverse separations components in the quarks wave 
function. We analyze the relation between transverse momentum 
components and transverse separations.  
Wave functions with large transverse momentum components 
can simultaneously reproduce the behavior of the Dirac form factor at 
large momentum transfer, and of the deep inelastic structure functions at 
Bjorken $x \rightarrow 1$. The presence of large momentum components
does not ensure, however, the dominance of small transverse distances
at large $x$.
We suggest that experiments measuring the attenuation of hadrons
in the nuclear medium, or the onset of color transparency, can 
provide an alternative source of information on generalized 
parton distributions, by mapping out the behavior of the transverse
components of the wave function.     
\end{abstract}

\maketitle

\section{Introduction}
A most intensively studied question in Quantum ChromoDynamics (QCD)
is the space-time structure of high energy exclusive reactions. 
In perturbative QCD (pQCD) approaches the hypothesis is made 
that elastic processes 
are dominated by the Fock space components of the hadronic 
wave function with the 
minimum number of quarks (anti-quarks). In the hard scattering
picture these are assumed to be located 
within a small relative transverse 
distance $ \lesssim 1/\sqrt{Q^2}$, 
$Q^2$ being the (high) four-momentum transfer squared in the 
reaction \cite{BroMul}. 
After the hard scattering takes place, 
a normal hadron is recovered.
The validation through experiments 
of this picture -- the dominance of the short
separation components --  
is however not straightforward, as witnessed for instance by the recent 
controversial interpretations of experiments sensitive to 
helicity selection rules \cite{Bro_rev,JP_01}.

It has been long surmised \cite{BroMul,RalPir} 
that performing (quasi)-elastic reactions off nuclear targets 
might ease this impasse. Nuclei can in fact 
function as both ``passive'' or ``active'' probes for the small partonic 
separation components. Small distances can be filtered at finite $Q^2$, 
by impeding the passage of large separations because
of the strong interactions occurring in a nucleus with $A >> 1$. 
In this case nuclei play an active role \cite{Kundu}. One can
also simply study the passage through the nuclear medium of the small 
separation component selected at asymptotically high $Q^2$, namely the phenomenon of
Color Transparency (CT). CT is attained
in principle when the ratio, $\displaystyle T_A(Q^2) = \sigma_A/A \sigma_h $, 
of the nuclear cross section to the one for scattering in free space, 
reaches the value of $1$ at some given high value of $Q^2$. 
From a practical point of view,  
current searches for CT
might appear to be in a stall as 
all experiments performed so far seem not to show 
either any systematics or any marked trend for the onset of 
this phenomenon \cite{exp_CT}. 
However, at the same time, none of the experiments
does contradict current predictions for CT.
A large body of theoretical work exists,  
that generated an active field in the 
last decade (see \cite{JaiPirRal,FMS,NNN} for extensive reviews).  
New observables have been proposed since the original 
measurements, 
such as the attenuation cross section \cite{Kundu}, 
the systematics of oscillations in pion production 
reactions \cite{JP_7}, or the sensitivity 
to the correlation length in exclusive vector meson 
production \cite{Boris}.
Based on these new hypotheses, more specific searches 
have been conducted, which in some cases seem to better identify 
the phenomenon \cite{HERMES_CT,Gao,Dutta,Ait}. 
Critical observations about the key assumptions underlying
the onset of CT have also recently emerged, a most disquieting one 
being the suggestion that the transverse size
of the exclusive process might not even be small due to a  
``suppression of the Sudakov suppression'' \cite{Hoyer}.

Whether or not a pQCD description of the hadrons holds at the 
$Q^2$ values presently available, it has now become imperative 
to investigate the basic question
of the existence and observability of small size hadronic 
configurations. 
In this paper we use as a theoretical tool 
Generalized Parton Distributions (GPDs) that,
as pointed out recently \cite{Burka}, can give us direct access to 
the transverse spatial extension of the hadronic components.
GPDs can, in fact, be shown to be 
diagonal in the impact parameter representation, and they can therefore
be interpreted as -- Fourier transformed in the transverse plane -- 
joint probability distributions of finding a parton   
with longitudinal momentum fraction $x$, and transverse separation from
the proton's center, $b$ \cite{Burka}. The original result, 
obtained for $\xi=0$, $\xi$ being the longitudinal momentum transfer 
fraction, was shown to hold also in the $\xi \neq 0$ case in \cite{Diehl_03}
(see also \cite{BelJiYuan} for a laboratory frame based description).  

The new insight provided by the impact parameter space interpretation  
allows one to make a connection between GPDs and 
the phenomena of CT and nuclear filtering.
Within this new framework, one can in fact address directly the question of
what transverse distances dominate the hard exclusive scattering 
cross section. 
So far, a quantitative knowledge of the size of the hadronic
configurations involved in a reaction 
has been in fact very elusive, or at the best, loosely inferred 
from experiments.

This paper is organized as follows: in Section \ref{def_sec} we 
define GPDs and show their connection, at leading order in $Q^2$, 
to both the impact parameter dependent parton distributions, and to
the intrinsic transverse momentum distributions.  
In Section \ref{mod_sec} we
evaluate the role of large transverse momentum components and, conversely,
of small interparton separations. A few quantities are calculated, namely
the GPDs at $\xi=0$, the Dirac form factor, and the the deep 
inelastic structure
function. For each of these quantities the impact of different behaviors
in the partons wave functions in both transverse momentum, and transverse
distance, is evaluated.    
In Section \ref{CT_sec} 
we calculate the ratio $T_A(Q^2)$ for electron nucleus scattering 
experiments in terms of GPDs, and we 
suggest a method for extracting the transverse 
interparton distance distribution. 
Finally, we draw our conclusions in Section \ref{con_sec}. 

\section{Some Old and New Definitions}
\label{def_sec}

GPDs were initially introduced  
\cite{Ji,Rad} in order to better describe in a partonic language 
the concept of orbital angular momentum carried by the nucleon's constituents. 
They can be accessed experimentally in exclusive hard scattering 
experiments. The most straightforward way has been to date 
Deeply Virtual Compton Scattering (DVCS), that has become the 
prototype for illustrating different aspects of GPDs.  
There exist a few reviews to date on this subject \cite{Goe,Die}. 
In this Section we summarize the properties of GPDs that are relevant 
for our discussion of exclusive scattering from nuclear targets, as 
{\it e.g.} probed in the reaction: $e A \rightarrow e^\prime p(A-1)$.
In particular, we define: 
\begin{itemize}
\item The Impact Parameter dependent Parton Distribution Function (IPPDF)
and its relation to the observables, $F_1$, 
the Dirac form factor, $\nu W_2$, the DIS structure function, and $H$,
the ``non-spin flip'' GPD.
\item The rms impact parameter, interparton separation, and radius for
the proton, in terms of both IPPDFs and GPDs.
\item The intrinsic transverse momentum distribution, and its relation to 
both the IPPDFs, and to $F_1$, $\nu W_2$, and $H$.
\end{itemize} 
The latter issue opens a new perspective on GPDs as ``tools'' to study 
transversity.  
  
\subsection{Generalities}
We consider the leading order in $Q^2$ (twist-2), 
where one can define four universal
GPDs, namely $H,E$ and $\widetilde{H}, \widetilde{E}$. The latter two are 
related to the quarks helicity distributions. In this paper we
focus on the unpolarized case described by $H, E$. 
The relevant kinematical variables are: $P$ and $P^\prime$, the initial 
and final nucleon's momenta in the exclusive process,  
$\bar{P}=(P+P^\prime)/2$, the average nucleon momentum, $k$ the 
active quark momentum, and $q$ the virtual photon momentum (Fig.1).
While the (upper) hard scattering part of the diagram varies in 
principle for different reactions, the information necessary to describe GPDs 
is contained in the (lower) soft part (the different types of matrix 
elements related to GPDs are described {\it e.g.} in \cite{Goe}).    
\footnote{We do not take into consideration initial state nuclear
medium modifications. These are part of a more 
general problem, and they will be addressed in a forthcoming paper 
\cite{LiuTan2}}
One can define four independent invariants using these variables:
\footnote{We use the notation: $a^\mu \equiv (a^0; {\bf a}_\perp, a^3)$, and
$a^{\pm} = (a^0 \pm a^3)$ }
$x=k^+/\bar{P}^+$, $\xi=-\Delta^+/2\bar{P}^+$, $Q^2 = -q^2$, and
$t \equiv \Delta^2$, where $\Delta=P-P^\prime$,
is the four-momentum transfer in the process. Differently
from inclusive Deep Inelastic Scattering (DIS), the four-momentum
transfer, $\Delta$, does not generally coincide
with the virtual photon four-momentum, $q$.    
By properly devising experimental measurements that
reproduce this situation, namely $q \neq \Delta$,
one has therefore the unique possibility of 
``zooming'' in on partonic components, due to the high resolution
provided by large $Q^2$ (in the multi-GeV region), at the same time
scanning the hadronic target's spatial distribution, through 
the dependence on the variable $\Delta$.   
We denote by $H_i$ and $E_i$ the GPDs for a quark of type $i$. $H_i$ 
and $E_i$ satisfy the following relations: 
\begin{subequations}
\begin{eqnarray}
\int_{-1}^1 dx H_i(x,\xi,t) & = & F_1^i(t), \\
\label{def1}
\int_{-1}^1 dx E_i(x,\xi,t) & = & F_2^i(t),
\end{eqnarray} 
\end{subequations}
independently from $\xi$ (see \cite{Goe,Die}). $F_{1(2)}^i$ are the Dirac and Pauli form 
factors for the quarks of type $i$ in the nucleon. They are readily 
obtained from the measured proton and neutron form factors by using 
isospin symmetry \cite{Goe}.
For $x \geq 0$, 
\begin{equation}
H_i(x,0,0) \equiv q_i(x),
\label{def2} 
\end{equation}
{\it i.e.} one recovers the Parton Distribution Functions (PDFs), $q_i(x)$, 
in the in the forward limit. 
$E$ does not have an equivalent connection to parton distributions.  

\subsection{Impact Parameter Space} 
GPDs can be related \cite{Burka} to the
Impact Parameter space dependent PDFs (IPPDFs) 
(see {\it e.g.} Ref.\cite{Soper}). 
An IPPDF is defined as the joint distribution:
$\displaystyle d n_i/dx d{\bf b} \equiv q_i(x,{\bf b})$ --
the number of partons of type $q$ 
with momentum fraction $x=k^+/P^+$, located at
a transverse distance ${\bf b}$ (${\bf b}$ is the impact parameter) 
from the center of $P^+$ of the system.
\footnote{All transverse variables are denoted here by bold-face characters.}
The latter is defined as
$\displaystyle {\bf R} = (\sum k_n^+ {\bf b}_n)/P^+$,  
where $x_n^\mu\equiv(x^0;{\bf x}\equiv {\bf b},x^3)$ is the space-time variable for the 
$n-th$ parton, and $\displaystyle \sum k_n^+ =P^+$. 
The usage of Light Cone (LC) variables is motivated by 
the fact that by viewing  $x^+$ as the time variable, $P^+, p_n^+$ as the 
total and partonic masses, respectively,  
and ${\bf R}$ as the center of mass of 
the system, $P^+ {\bf R}$ is the generator of longitudinal boosts that 
leave the physics on the $x^+=0$ surface invariant.
An intuitive connection with non-relativistic logics is 
naturally provided. 
The interpretation of 
observables, {\it e.g.}  structure functions,  in terms of
parton density distributions relies on the validity of the Impulse
Approximation (IA) whereby the hard scattering happens instantaneously
in $x^+$.
We work in this hypothesis by assuming that it is satisfied in 
the kinematical limit for the coherence length of the virtual photon given by:
$x^+ \ll L \approx 1/2 M_N x_{Bj}$. This amounts to considering only 
large values of $x_{Bj}$. 
At low $x_{Bj}$, one has to consider both IA-type 
scattering and of processes including 
photon splitting into quark-antiquark pairs before the nucleon where it has
been recently shown that FSI might hamper the interpretation of the deep inelastic
measurements in terms of
probability distributions \cite{BHPS}. This is part of 
a wider set of problems that transcend the goals of this paper.  
In particular, it will be important to study
in the future whether parton interactions (or off-shellness) 
affect the transverse distributions
that we address in this paper, as this might complicate the procedure 
of extracting holographic images of the proton as proposed 
in \cite{femto,BelJiYuan}. 

By definition \cite{Soper}, The DIS structure function, 
$\nu W_2(x)$, and the form factor, 
$F_1(Q^2)$, can be written in terms of the IPPDFs respectively, as:
\begin{subequations}
\begin{eqnarray}
\nu W_2(x) & = & \sum_i e_i^2 \, x \int 
d ^2 {\bf b} \, q_i(x,{\bf b}) 
\label{sf} \\
F_1(\Delta) & =  & \sum_i e_i \int d ^2 {\bf b} \, 
e^{i {\bf \Delta} \cdot {\bf b}} \int _0^1 dx \, q_i(x,{\bf b}),
\label{ff}
\end{eqnarray}  
\end{subequations}
where $e_i$ is the quark's charge, and we have disregarded the 
$Q^2$ dependence of $\nu W_2$. Notice that ${\bf \Delta}$ is the 
conjugate variable of ${\bf b}$.

\subsubsection{{\bf b} distribution}
In \cite{Burka}, the observation was made that for a purely 
transverse four momentum transfer, namely for 
$\Delta \equiv (\Delta_0=0; {\bf \Delta}, \Delta_3=0)$ and $\xi=0$,
$H_i(x,0,-{\bf \Delta}^2)$, 
and $q_i(x,{\bf b})$ can be 
related as follows:
\begin{subequations}
\begin{eqnarray}
q_i(x,{\bf b}) & = & \int \frac{d^2 {\bf \Delta}}{(2 \pi)^2} \,
e^{-i {\bf b} \cdot {\bf \Delta}} H_i(x,0,-{\bf \Delta}^2)
\label{bdis1}  \\
H_i(x,0,-{\bf \Delta}^2) & = & \int d^2 {\bf b} \,
e^{i {\bf b} \cdot {\bf \Delta}} q_i(x,{\bf b})
\label{bdis2}  
\end{eqnarray}
\end{subequations}
Since $q_i(x,{\bf b})$ satisfies positivity constraints and it can
be interpreted as a probability distribution, $H_i(x,0,-{\bf \Delta}^2)$
is also interpreted as a probability distribution, namely the Fourier 
transformed joint probability distribution 
of finding a parton $i$ in the proton with longitudinal momentum fraction 
$x$, at the transverse position ${\bf b}$, with respect to the
center of momentum of the nucleon. Since in what follows we will consider only 
the case $\xi=0$, we define: 
$H_i(x,0,-{\bf \Delta}^2) = H_i(x,{\bf \Delta})$. 

The root-mean-square (rms) distance of the configuration with 
momentum fraction $x$, 
from the center of the proton is defined as the square root of:
\begin{equation}
\displaystyle
\langle \, {\bf b}^2(x) \, \rangle = {\cal N} \int d^2 {\bf b} \; q(x,{\bf b})
\, {\bf b}^2,
\label{b}
\end{equation}
${\cal N}$ being a normalization factor, {\it i.e.}  
$\displaystyle {\cal N}^{-1} = \int  d^2 {\bf b} \; q(x,{\bf b})$. 
Similar relations to the ones in Eqs.(\ref{bdis1},\ref{bdis2}), were obtained
for $\xi \neq 0$ in \cite{Diehl_03}.
An alternative approach in terms of Wigner distributions, 
formulated in a 
covariant model that allows one in principle to perform calculations in the 
laboratory frame was given in \cite{BelJiYuan}. 
The role of the variable $\xi$ in the context of
a covariant approach, is an important subject {\it per se}, especially in 
reference to Initial State Interactions (ISI) in nuclei, since
it impacts the particles' off-shellness. 
We address this subject in detail in \cite{LiuTan2}. 

\subsubsection{{\bf y} distribution and hadronic radius}
One can also define the Fourier transform with respect to the variable 
${\bf y} = {\bf b}/(1-x)$, the transverse position of parton
$i$ with respect to the center of momentum of the system of 
spectator partons \cite{Soper}:
\begin{equation}
{\cal H}_i(x,\widetilde{\bf {k}}) = \int d ^2{\bf b} \, e^{i \widetilde{{\bf k}} \cdot {\bf y}} 
 \, q(x,{\bf b}) 
\label{soper_var}
\end{equation}
${\cal H}$ can be obtained from $H$ by rescaling the functions 
arguments as follows:
\begin{equation}
H_i(x,{\bf \Delta}) \equiv  {\cal H}_i(x, {\bf \Delta}(1-x) )
\end{equation} 
Since ${\bf y}$ measures the interparton separation, 
it is useful to define also the inverse Fourier transform: 
\begin{equation}
\widetilde{{\cal H}}_i(x,{\bf y}) = 
\int d {\bf k}^\prime \, e^{-i {\bf \widetilde{k}} \cdot {\bf y}} 
 \, {\cal H}_i(x,{\bf \widetilde{k}}) \equiv  q\left(x,{\bf y}(1-x)\right).
\label{soper_var2}
\end{equation}
The variables 
$\widetilde{{\bf k}} \equiv {\bf \Delta} (1-x)$ and ${\bf y}$ are Fourier conjugates,
as well as ${\bf \Delta}$ and ${\bf b} \equiv {\bf y}(1-x)$. 
$\widetilde{{\cal H}}(x,{\bf y})$ is the probability of finding a parton with  
momentum fraction $x$, at a transverse distance ${\bf y}$ from the center of 
momentum of the spectator partons.
The rms transverse distance between partonic 
configurations with momentum fractions $x$ is obtained as the square root of:
\begin{equation}
\langle \, {\bf y}^2(x) \, \rangle = {\cal N} \int d^2 {\bf b} \,  q(x,{\bf b({\bf y})})\, {\bf y}^2 ,
\label{y}
\end{equation}
{\it i.e.} from the same observable $H$ used to obtain 
$\langle {\bf b}^2 \rangle$. ${\cal N}$ is defined analogously to Eq.(\ref{b}). 

The {\em radius} of the system of partons, which is the useful quantity 
for our CT study, is:
\begin{equation}
\displaystyle \, \langle {\bf r}^2(x) \, \rangle ^{1/2} = 
MAX\left\{ \langle {\bf b}^2(x) \rangle ^{1/2}, \langle {\bf b}^2(x) \rangle ^{1/2}
\frac{x}{1-x}   \right\}
\label{r}
\end{equation} 
This definition coincides with the one given in \cite{Burka} 
for $x\rightarrow 1$. In Fig.\ref{fig2} we sketch a cartoon of the
different transverse distances discussed so far. The figure 
shows three different scenarios for the behavior
of such distances with increasing $x$, 
corresponding to the models discussed in the following Sections. 
\subsubsection{Intrinsic transverse momentum 
distribution and spectral function}
\label{intkperp_sec}  
The function ${\cal H}_i(x,\widetilde{{\bf k}})$ defined in the previous subsection,
although being related through a Fourier transform to the interparton 
separation, ${\bf y}$, is conceptually different from the 
``intrinsic transverse momentum'' distribution.
 
Although we started from writing the structure function 
and the form factor in terms of IPPDFs (Eqs.(\ref{sf}) and (\ref{ff}), 
respectively),  
their definition is usually given
in terms of a LC wave function, $\phi(x, {\bf k})$ as \cite{Lep}:
\begin{subequations}
\begin{eqnarray}
\nu W_2(x) & = & \sum_i e_i^2 x \int d ^2{\bf k} \, \, f(x,{\bf k}) 
\label{sf_kperp} \\
F({\bf \Delta}) & =  & \sum_i e_i \int d ^2 {\bf k} 
\int _0^1 dx f(x,{\bf k},{\bf k}+(1-x){\bf \Delta}),
\label{ff_kperp}
\end{eqnarray}  
\end{subequations}
where:
\begin{equation}
f(x,{\bf k},{\bf k^\prime}) = \phi^*(x, {\bf k}) \phi(x, {\bf k^\prime}), 
\end{equation}
with ${\bf k}^\prime \equiv {\bf k}+(1-x){\bf \Delta}$, is a non-diagonal intrinsic
momentum distribution. The diagonal term can be written as:
\begin{equation}
f(x,{\bf k}) = \mid \phi(x,{\bf k}) \mid^2. 
\end{equation}

By comparing Eqs.(\ref{sf_kperp}, \ref{ff_kperp}), with 
Eqs.(\ref{sf}, \ref{ff}), we find that
the relation between the transverse momentum and the transverse separation 
of quarks inside a hadron is obtained through a non-diagonal distribution 
in transverse coordinate space, $q(x,{\bf b}, {\bf b}^\prime)$:
\begin{equation}
f(x,{\bf k}) = \int d^2 {\bf b} \int d^2 {\bf b}^\prime \, 
e^{i {\bf k} \cdot ({\bf b} - {\bf b}^\prime)} \, q(x,{\bf b}, {\bf b}^\prime),
\label{intrinsic_kt}
\end{equation}
where we define:  
\begin{subequations} 
\begin{eqnarray}q(x,{\bf b}, {\bf b}^\prime) & = & \Psi^*(x,{\bf b}^\prime) \Psi(x,{\bf b}), \\
\label{nondiag_b}
q(x,{\bf b}, {\bf b}) & = & \mid \Psi(x,{\bf b}) \mid ^2 \equiv q(x,{\bf b}),
\end{eqnarray}
\end{subequations}
$\Psi(x, {\bf b})$ being a wave function in the mixed $x$ and ${\bf b}$ 
representation, related to the LC
wave function  by a transverse Fourier transform. 
This result can be also implemented in studies of processes directly
sensitive to ${\bf k}$.
It is useful for this purpose to consider the average transverse 
momentum squared, defined as:
\begin{equation}
\displaystyle
\langle \, {\bf k}^2(x) \, \rangle = {\cal N} \int d^2 {\bf k} \; f(x,{\bf k})
\, {\bf k}^2,
\label{k}
\end{equation}
with $\displaystyle {\cal N}^{-1} = \int d^2 {\bf k} \; f(x,{\bf k})$.
In Section \ref{mod_sec} we present an initial exploratory
study of the relative roles of the intrinsic transverse momentum and  
the transverse separation using different types of wave functions.

Finally, it should be mentioned that Eq.(\ref{sf_kperp}) corresponds to 
a projection obtained from the partons covariant spectral function, $S(k)$ 
\cite{GroLiu,KPW,Mul}:
\begin{eqnarray}
\displaystyle \nu W_2(x) & = & \sum_i e_i^2 x \int d ^4 k \, \, 
S(k) \, \delta\left(x-\frac{k^+}{P^+} \right)  \nonumber \\
\displaystyle & = & \sum_i e_i^2 x \int d^2 {\bf k} \int d P_X^2  \, S(k),
\label{spefunc}
\end{eqnarray} 
where the integration over $k^-$ in the last line 
is replaced by an integration over the invariant 
mass of the spectator system, $P_X^2=(P-k)^2$ ($k^-$ and $P_X^2$ are related by).  
The main reason for introducing the spectral function along with the intrinsic
momentum distributions is that it encodes information about the energy
and momentum of the spectator system. The spectral function can be defined in fact 
as the joint probability distribution of finding a parton with LC components 
$k^+$ and ${\bf k}$, leaving the spectator system with a mass $M_X^2 \equiv P_X^2$.   
The usage of the spectral function is most important in problems where the particles' 
off-shellness/transversity plays a role, a typical example being deep
inelastic scattering from nuclei \cite{LiuTan2,GroLiu,Mul,KPW}. 
In the following 
Sections we show a model of GPDs that implements a function $S(k)$
within relativistic IA.  

\section{Models for Impact Parameter Dependent Parton Distribution
Functions}
\label{mod_sec}
We evaluate the functions defined in the previous Section at leading order in 
$Q^2$, using a two component, or spectator, model (Fig.1b). 
The main assumptions of the model are: {\it i)} the validity of the 
hand-bag diagram; {\it ii)} a description of dynamics in terms of a 
nucleon-quark-spectator system vertex. The minimal Fock component
is described by a quark-diquark vertex function.
We describe the active quark as off-shell and the diquark as on-shell, with
mass $M_X = \sqrt{P_X^2}$. 

The spectator model is both widely used and well tested in hadronic physics. 
In particular, in \cite{GroLiu} it was used to describe 
DIS in nuclei; in \cite{Mul} 
parametrizations for both the polarized and unpolarized proton and 
neutron structure functions were given; more recently  
it was applied to DVCS \cite{Kroll,Lon}. 
Our usage of the model other than for 
practical reasons, is motivated by the clear handling 
of the interplay between the particles' off-shellness and their energy 
and transverse momentum. This turns out to be an important property when 
describing the deep inelastic structure of nuclei \cite{LiuTan2}.
By considering the components in Fig.1b, one has
\begin{subequations}
\begin{eqnarray}
k^2  & = &  k^+ k^- - {\bf k}^2 = \nonumber \\
 & = & (xP^+) \, 
\left[\frac{M^2}{P^+} - \frac{M^2_X + {\bf k}^2}{(1-x)P^+} \right] 
- {\bf k}^2  \\
\label{vir}
k^{\prime 2} & = &  k^{\prime +} k^{\prime -} - ({\bf k}+ {\bf \Delta})^2 = 
\nonumber \\ 
 & = & (xP^+) \, 
\left[\frac{M^2+{\bf \Delta}^2}{P^+} - \frac{M^2_X + {\bf k}^2}{(1-x)P^+} 
\right] 
- ({\bf k}+ {\bf \Delta})^2 
\label{vir_prime}
\end{eqnarray}
\end{subequations}  
where ${\bf \Delta}^2 \equiv -t$;  
$M$ and $M_X$ are the proton and the diquark masses, respectively.   
Hence, for the denominators in Fig.1b:
\begin{equation}
D(x,{\bf k})  =  {\cal M}_X^2 \, x - \frac{{\bf k}^2}{1-x}, 
\end{equation}
where we defined ${\cal M }_X^2 = M^2 \,x - M_X^2/(1-x) - m^2/x$,
$m$ being the struck quark mass.
We consider the remnant diquark system to be a scalar. 
The form of the vertex function depends on the assumptions
made in treating the quark's spin. Since we are ultimately interested
in determining the radius of the proton as extracted by using the spin
averaged quantity $H$, we assume that the vertex function is a 
scalar in Dirac space described by a function 
$g(k^2)$.  
The two-component wave function becomes:
\begin{equation}
\phi(x,{\bf k})  =  \frac{g(k^2)} {D(x,{\bf k})}. 
\label{wfunction_k}
\end{equation}
Notice that, because of the relation between the struck parton's virtuality
and transverse momentum (Eq.(\ref{vir})), one can write $\phi$ as
a function of the invariants: $k^2$, $x$, $M_X^2$. 
One can therefore identify the relativistic spectral function in 
Eq.(\ref{spefunc}) with:
\begin{equation}    
S(k) = \mid \phi(k^2,M_X^2,x) \mid^2.
\end{equation}

\subsection{GPD, Structure Function, and Form Factor}
The GPD, the structure function, and the form factor can be evaluated
in terms of the function $\phi$ in Eq.(\ref{wfunction_k}), according to
the definitions in Section \ref{intkperp_sec}.
One has:
\begin{subequations}
\begin{eqnarray}
H(x,{\bf \Delta}) & = &\int d^2 {\bf k} \, 
\phi^*(x,{\bf k}) \phi(x,{\bf k}+(1-x){\bf \Delta}) 
\label{gpd3}
\\
\nu W_2(x) & = & \int d ^2{\bf k} \, 
\mid \phi(x,{\bf k}) \mid^2
\label{sf3}
 \\
F_1({\bf \Delta}^2) & =  & \int d ^2 {\bf k} \, 
\int _0^1 dx \, \phi^*(x,{\bf k}) \phi(x,{\bf k}+(1-x){\bf \Delta}),
\label{ff3}
\end{eqnarray}  
\end{subequations}
Notice that we incorporated in the function 
$\phi$, a factor $\displaystyle \sqrt{x/(1-x)}$, that accounts for both the initial
flux, and the phase space for the process in Fig.1.  
Moreover, we drop the index $i$ in $H$, since we fit our functions
directly to the proton valence structure function and form factor, without
distinguishing between $u$ and $d$ quarks contributions.  
For the vertex function we used the form: 
\begin{equation}
g(k^2) = g \, \frac{\Lambda^2-m^2}{k^2- \Lambda^2},
\label{mon}
\end{equation}
where $g$ is fixed by the normalization, and $\Lambda$ is a cut-off
parameter. The values of $\Lambda$, $m$, and $M_X$ are determined by 
fitting both the large four-momentum behavior of the form 
factor, and the shape of the DIS structure function at large $Q^2$.
The latter is determined in our model calculation at very low 
$Q^2$ ($Q^2 < 1$ GeV$^2$), where the structure function is given only
by its valence component, the sea quarks and gluons being 
generated through perturbative evolution (see {\it e.g.} \cite{GRV_OLD}).     
The values of the parameters are: $m= 0.3$ GeV, $M_X=1.100$ GeV,
$\Lambda = 0.73 $ GeV.  
In Figures 3, 4, 5, and 6 we present our results obtained using
Eqs.(\ref{wfunction_k},\ref{gpd3},\ref{sf3},\ref{ff3},\ref{mon}).

Fig.3 shows the quantity: $\displaystyle \mid \Phi({\bf k}) \mid^2$, where
\begin{equation} 
\displaystyle \Phi({\bf k}) = \int_0^1 dx \, \phi(x,{\bf k}) .
\label{funck}
\end{equation}
Also shown for comparison are results 
obtained using the parametrizations from 
Refs.\cite{Rad98,Stoler,Bur_new}.
The wavefunction in Eq.(\ref{wfunction_k}), 
has a similar behavior to the parametrization used in \cite{Stoler},
in that it displays high momentum components with a $\propto 1/{\bf k}^4$ 
behavior, although in a slightly larger amount.
In order to quantify the ``hardness'' of the distributions, in Table 
\ref{table_k2} we show  
the average transverse momentum values obtained in each model.
In the Table we also display the contributions to 
$\langle {\bf k}^2 \rangle$, from the large momentum components 
calculated by setting the lower limit of integration 
in $\mid {\bf k} \mid$ equal to
$1$ GeV and to $2$ GeV, respectively.
These values reflect the behavior of the curves in Fig.\ref{fig3}, 
in that the distribution in Ref.\cite{Rad} is truly of the ``soft'' type;
the ones in Ref.\cite{Stoler} and in this paper are characterized by a long
high momentum ``tail''; finally, we show results obtained with 
parametrizations of the type proposed
in Ref.\cite{Bur_new}, where by accounting for gluonic interactions
through the introduction of an effective gluonic mass in the initial
two component model of \cite{Rad98}, a further suppression in the 
large $x$ behavior is found, of the type 
$\approx \displaystyle \exp[(1-x)^n \Delta^2]$, with $n \geq 2$. 
We find that for $n=2,3$ the distributions obtained following the idea of 
Ref.\cite{Bur_new}, are neither entirely ``soft'' nor ``hard'', but they
display larger ``intermediate'' momentum values, hence they give rise to
a large $\langle {\bf k} \rangle ^2$. 

\begin{table}
\caption{Contribution of large transverse momentum components from different
GPD models: 
values of $\langle k^2 \rangle$ (first column); percentage contribution 
to $\langle k^2 \rangle$ from $k > 1$ GeV components (second column), 
and from $k > 2$ GeV components (third column).} 
\begin{tabular}{|c|c|c|c|}
\hline 
  & $(\langle k^2 \rangle^{1/2}  \, {\rm MeV})$ & $\%$ value for $k_{min}= 1$ (GeV) & $\%$ value for $k_{min}=2$ (GeV)  \\
\hline
\hline
This paper & 360 & 14 & 1.5   \\
Radyushkin '98 \cite{Rad98} & 290 & 0.1  & 0  \\
Stoler '02 \cite{Stoler} & 286 & 2.0 & 1.0  \\
Burkardt '04 \cite{Bur_new} & 415 & 15 & 6 $\times $ 10$^{-2}$ \\
\hline 
\end{tabular}
\label{table_k2}
\end{table}

The function $H(x,{\bf \Delta})$ is shown 
in Fig.4, plotted
vs. ${\bf \Delta}$ for different values of $x$. A small ($x=0.07$), 
an intermediate ($x=0.36$) and a large ($x=0.88$)
value of $x$ are indicated in the figure, from which one can 
also deduce that the 
value of $H(x,0)$ traces the expected behavior for the valence structure 
function (namely the curve at $x=0.07$ has a much lower value for
${\bf \Delta} \rightarrow 0$ than the curve at $x=0.36$). Values of $x \gtrsim 0.9$ decrease one more order of magnitude
and they are not shown in the figure. 

In Fig.5 we show the form factor $F_1(\Delta^2)$. 
This is compared to calculations using the models of
Ref.\cite{Rad}, Ref.\cite{Stoler}, and to the dipole 
model: $F_1(\Delta^2) = {\cal F} 1/(1+\Delta^2/0.71)^2$,
with $\displaystyle {\cal F} = (\Delta^2/4M^2 \mu_P+1)/(1+\Delta^2/4M^2)$. 
We find, in agreement also with Ref.\cite{Stoler}, that the large momentum 
components of the wave function are responsible for the large  
$\Delta^2$ behavior of the form factor. Such components  
are effectively comparable to the hard contribution from pQCD calculations 
\cite{Kroll}. 
Although they constitute a small fraction of the 
average transverse momentum as shown in Table I, the contribution of 
large ${\bf k}$ is much larger in the nucleon form factor.
This is shown in Fig.6, where we plotted the ratio: 
$F_1({\bf \Delta}, {\bf k}_{max})/F_1({\bf \Delta}, {\bf k}_{max}=\infty)$. 
Only $\approx 70 \%$ of the 
form factor is reproduced by stopping the integration at $k_{max}=0.7$ GeV, 
by using a wave function with hard components, 
whereas the ratio shown in the figure is already saturated in 
models based on a Gaussian behavior. 

Similar conclusions can be drawn by investigating the large $x$ behavior 
of the deep inelastic structure function, Eq.(\ref{sf3}).

\subsection{Impact Parameter Space Parton Distributions}
Next, we study how the behavior 
of the wave function with transverse momentum, ${\bf k}$, reflects 
on the transverse space distribution, $q(x,{\bf b})$. 
This is obtained as a Fourier transform of the function $H(x,{\bf \Delta})$ in the transverse plane: 
\footnote{For a circularly symmetric 
function of $\Delta$ this is a Hankel transform}
\begin{equation}
q(x,b) = \frac{1}{2 \pi} \int_0^\infty d \Delta \; \Delta \; J_0(b\Delta) \, H(x,\Delta) 
\label{hankel} 
\end{equation}
where $\Delta = \mid {\bf \Delta} \mid$ and $b = \mid {\bf b} \mid$, and
$J_0$ is a zero-th order Bessel function of the first kind.
In order to evaluate $q(x,b)$ with 
adequate accuracy, we parametrized $H(x,\Delta)$ as:
\begin{equation}
H(x,\Delta) = \frac{p_1(x)}{(p_2(x) + p_3(x) \Delta^2 + p_4(x) \Delta)^2} . 
\label{H_par}
\end{equation}
The coefficients $p_i$, $i=1,4$ are given in 
the Appendix.
The Hankel transform of $H$ has the following analytic form:
\begin{equation}
q(x,b) = \frac{1}{2} \, b \frac{p_1(x) p_2(x)}{\sqrt{p_3(x)/p_2(x)}} 
K_1(\sqrt{p_2(x)/p_3(x)} \, b),
\label{num_qxb}
\end{equation} 
where $K_1$ is a modified Bessel function of the second kind, and
we set $p_4 =0$ (a more complicated, though similar, 
analytical form is obtained for $p_4 \neq 0$, which we do not
display for ease of presentation).
$q(x,b)$ is shown in Fig.7, plotted as a function 
of $b$ for varying $x$. 

Using Eq.(\ref{num_qxb}) one can calculate 
the radius, the interquark separation, 
and the rms distance from the center of the hadron, 
Eqs.(\ref{b},\ref{y},\ref{r}). These quantities are shown
in Fig.8. 
In Fig.8(a) results for $\langle b^2 \rangle^{1/2}$ from several models for the GPDs 
are compared. The difference in slope as $x \rightarrow 1$ is clearly visible.
In Fig.8(b) we show our results for $\langle b^2 \rangle^{1/2}$, $\langle y^2 \rangle^{1/2}$,
$\langle r^2 \rangle^{1/2}$.
One can see that the both rms radius, and the interparton separation 
do not vanish in the $x\rightarrow 1$ limit;  
rather, they tend to a finite 
(small) value of $\approx 2.5$ GeV$^{-1}$ ($ \approx 0.5$ fm).
In Fig.9 we plot the average transverse momentum squared of a 
quark in the proton, Eq.(\ref{k}). There is a marked difference in the behavior
of this quantity for the ``soft'' and ``hard'' type wave functions. The wave
function proposed in \cite{Bur_new} gives manifestly divergent results 
for large $x$, although, as one can see from Table I, the integrated 
value of $\langle {\bf k}^2 \rangle$ is finite. Our results are consistent
with previous findings using the diquark/spectator model \cite{Mul}.        

What components of the wave function are responsible for this behavior?
In Fig.10 we give a partial answer to this question by plotting
the contribution to both $b$ (upper panel) and $r$ (lower panel) of the
short distance components. This is obtained by stopping the integration 
over $b$ in the definition of the rms quantities at decreasing 
values of the upper limit: 
$2$, $1$, and $0.5$ GeV$^{-1}$, respectively.
It is clear that the short distance part of the
wave function (or the large $k$ components) largely determine the size of the radius
at large $x$. This is particularly evident from the lower panel of Fig.10,
where results are presented using the ratio: 
$\langle r^2(b_{max}) \rangle/\langle r^2(\infty) \rangle$. 
While models including a hard component saturate this ratio for 
$b<0.5$ GeV$^{-1}$ already at $x=0.8$, models using a ``soft'' wave function
of the type of Ref.\cite{Rad98} do not show a tendency to saturation until
very large $x$.  
These results open up a somewhat puzzling situation:   
In order to describe the form factor at large $Q^2$, one needs 
high $k$ components in the wave function. This is both a prediction
of pQCD based calculations \cite{Bro_rev,JP_01}, and it
can be also shown phenomenologically (Fig.5) (see also Ref.\cite{Stoler}).
High $k$ components are also responsible for  
a decrease in the hadronic transverse size as $x$ 
increases (Figs.8, 9 and 10).
However, one can envisage a model such as the one utilized in this paper 
that has the amount of high momentum components 
necessary to reproduce the Dirac form factor, and that 
at the same time, predicts a small but not vanishing transverse 
size for hadronic configurations at large $x$.

In concluding this Section, we reiterate that the usage of GPDs for 
determining the spatial structure of hadrons through $q(x,b)$, 
has been so far largely theoretical, since these quantities are only indirectly 
deduced from a variety of delicate  
measurements (\cite{dvcs_HERMES,dvcs_Jlab}).   
In the following we suggest that an additional source of information could be 
given by studies of Color Transparency and related phenomena such as 
Nuclear Filtering. These are in fact directly sensitive to the size of 
hadronic components.

\section{Exclusive Scattering in the Nuclear Medium}
\label{CT_sec}
In the previous Sections we discussed the relative importance 
of the soft and hard components of the light cone
wave functions up to four momentum transfer $Q^2= 25-30 $ GeV$^2$.
Plausible scenarios confirm the idea that 
$ \approx 30 \%$ of the form factor is given by the hard component of 
the wave function at four momentum transfer $Q^2= 25-30 $ GeV$^2$ 
(for consistency with
the general notation we switch from $\Delta^2$ to $Q^2$ in this Section).
This result is indeed correlated with the dominance of small transverse 
sizes of 
hadronic configurations through a Fourier transform, however 
not straightforwardly as
demonstrated by the different behaviors of the GPDs with 
$x$ and $\Delta$. As shown in Fig.11 where the average value of
$x$, given by:
\begin{equation}
x_{ave}(\Delta) = \frac{\int_0^1 dx \, x \, H(x,\Delta)}{\int_0^1 dx H(x,\Delta)},
\end{equation}
is plotted vs. $\Delta \equiv \sqrt{Q^2}$, 
the model in Ref.\cite{Rad98} is governed at large $Q^2$ 
by small $b$ 
{\it as well as} by large $x$ components
(although its prediction for the form factor is substantially 
lower than the data, as shown in Fig.\ref{fig5}, and the 
radius predicted for hadronic configurations diverges at large $x$).
The approach based on the diquark model yields a form factor 
that is dominated by small $b$ components 
but not exclusively by $x \rightarrow 1$
configurations. This behavior which is further clarified by
the comparison of $H(x,\Delta)$ for the two cases in Fig.12, 
is at the origin of the rather 
flat dependence with $x$ of the transverse radius shown in Fig.8.   
Also, it appears that the model of Ref.\cite{Bur_new}, 
approximately including the effect of a gluonic interaction,
fixes the problem of an unphysically large radius while keeping
a large $x$ behavior as the one in Ref.\cite{Rad98}, at the expense 
of introducing a large $k$ in the wavefunction (Fig.\ref{fig9}). 

In summary, while it can be challenging to unambiguously disentangle the 
amount and nature of hard components responsible for the large $Q^2$ behavior
of the hadronic form factors, one might 
gain a better insight by requiring models to simultaneously 
describe the hadrons transverse spatial distribution, and in particular
the possible onset of small transverse configurations.   
The diquark model presented in this paper seems to 
provide a better description than other models similarly  based on the 
handbag diagram. 

If configurations with small radii indeed exist,
they can be isolated in principle by performing CT and/or nuclear filtering
type experiments. 
CT has been so far investigated in a variety of reactions including 
proton-hadron, $\gamma$-hadron, and electron-proton
scattering \cite{exp_CT,HERMES_CT,Gao,Dutta}.  Here we consider  
exclusive electron-proton scattering, where oscillations due to the 
interference of perturbative and non-perturbative contributions possibly 
present in the proton-proton and $\gamma$-proton exclusive cross 
sections, do not occur. Our result is however quite general and 
can be easily extended to other type of reactions.

Using the new perspective of GPDs in impact parameter space, we can now 
take a radical turn and devise a ``minimal'' approach
that reduces the (nuclear) model dependence of the problem
(an approach complementary to ours was recently suggested 
in \cite{BurMil}, where 
the connection between CT and 
GPDs was explored in relation to 
diffractive dissociation of pions).  
As widely underlined in the 
literature, the prerequisites for the onset of CT,
assuming that the scattering takes place off a 
small size configuration, are that: {\it i)} the interaction of the hadron
in the nuclear medium is reduced due to the decreased gluon radiation
from a small transverse size color dipole \cite{Nus,Bertsch}; 
{\it ii)} the small size configuration, not being stationary, will in 
principle evolve with time into a larger configuration. It however remains
sufficiently small during the time it crosses the nucleus.    
Point {\it ii)} has probably been the most controversial one, although 
recent criticism to {\it i)} has also emerged \cite{Hoyer}. 
Another point, not extensively addressed in the literature (see however
\cite{FMS}), is that
the hadron suffers Initial State Interactions (ISI), or that in other
words, the form factor for an off-shell nucleon is in principle
different from the on-shell one.

We can now break down the problem in several steps:
We first assume that the scattering happens with an ``unmodified'' 
proton in the nuclear medium. We introduce a nuclear filter for the 
large transverse size components, 
by multiplying the IPPDF, $q(x,b)$, by a square function:
\[ \Pi(b) = \left \{  \begin{array}{c} 
1  \; \; b <    b_{max}(A) \\
0 \; \;  b \geq b_{max}(A) \end{array} \right.   \]
where $b_{max}(A)$ defines the size of 
the filter. This affects the GPD as:
\begin{equation}
H_A(Q^2) = \int_0^{b_{max}(A)} db \, b \, q(x,b) J_0(b\Delta),
\label{transp1}
\end{equation}
with $\Delta = \sqrt{Q^2}$. 
The transparency ratio for $(e, e^\prime p)$ type reactions 
becomes 
\begin{equation}
\displaystyle T_A(Q^2) =  
\frac{\left[ \int_0^1 dx H_A(x,\Delta) \right] ^2}{\left[ \int_0^1 dx H(x,\Delta) \right]^2}, 
\end{equation}
where $H$ defined as in Eq.(\ref{bdis2}) represents scattering in free 
space, {\it i.e.} it is calculated with 
$b_{max}=\infty$.
Based on this result, one can fit the available data, using 
different distributions $q(x,b)$, and
varying the parameter $b_{max}$.   
In Fig.\ref{fig13} we show a few possible scenarios based on a simplified
analytical model for $q(x,b)$. We take 
$\displaystyle q(x,b) = A(x) \exp(-\alpha(x) \, b)$, 
where $\alpha(x)$ is taken with two different dependencies:
{\bf (a)} $\alpha \propto (1-x)/x$, as in Ref.\cite{Rad98}; {\bf (b)} 
$\alpha \propto (1-x)^2$, as in Ref.\cite{Bur_new}.  
The effect of the filter is to produce both damping and oscillations
in $H_A$. In Fig.\ref{fig13} we show for illustration, the ratio
$R = H_A(x,\Delta)/H(x,\Delta)$ plotted vs. $\Delta$ for two different values
of $x$, in case {\bf (a)} and {\bf (b)}, and
for different values of the filter size. Our analytical calculation
displays explicitly the damping and oscillations as:
\begin{equation}
R = 1- \exp(-\alpha \, b_{max})
\left( \alpha \, b_{max} J_0(\Delta b_{max}) + \cos(\Delta b_{max}) \right). 
\end{equation} 
Although this result does not correspond to a completely 
realistic situation, 
it allows us to understand for varying $x$, the different effects due 
to variations 
in the size of $\alpha$, which in turn is a feature of GPDs 
from different models. Oscillations are clearly more   
pronounced at low $x$. When $b_{max}$ is large, of the order of $5$ 
GeV$^{-1}$ ($\approx 1$ fm), transparency is attained. When $b_{max}$ is 
small ($1$ GeV$^{-1}$, $\approx 0.2$ fm), the medium is no longer 
transparent to the type of distributions considered, although distribution
{\bf (a)} giving a larger radius, is more suppressed than {\bf (b)} 
at low $\Delta$.  

The results of the first step of the analysis would give 
joint information
on possible sizes of the filter, defined as 
$y_{max} \equiv b_{max}/(1-x)$, and on the behavior of 
the IPPDFs, {\it i.e.} 
on the radius of hadronic configurations, $\langle {\bf r}^2 \rangle^{1/2}$.   

This procedure, systematically applied 
to a sufficiently large body of data including both 
existing and planned measurements \cite{exp_CT,Gao,Dutta,Jlab_01_107},   
gives a much more direct test of the transverse sizes involved.
Once this basic information is known from measurements -- and not
inferred from theoretical scenarios -- 
one would be able to introduce more sophisticated modeling 
of {\it e.g.} rescattering and ISI.
A somewhat similar point of view was taken in a pre-GPD context
in Ref.\cite{JP_8} and applied to the data in \cite{oneill,garrow},
where fits were performed based on a parameter, 
$p=n_A \, R_A \, \sigma_{eff}$, 
with $n_A$ the nuclear density, $R_A$ the nuclear radius, and 
$\sigma_{eff}$ the effective hadron-hadron interaction cross section.


\section{Conclusions and Outlook}
\label{con_sec}
We presented a study using the new theoretical insight 
provided by GPDs, of the interplay between the transverse 
variables in both momentum and coordinate space, ${\bf k}$ and ${\bf b}$ 
respectively, and of the longitudinal
momentum fraction, $x$, in the proton form factor, $F_1$. 
Our study was aimed at establishing what type of components 
-- hard vs. soft, small transverse distance vs. large  -- 
dominate the form factor
based on different hypotheses for the GPDs used 
to describe $F_1$.

We examined three types of distributions (Figs.\ref{fig3} and \ref{fig4}): 

\vspace{0.3cm}
\noindent
{\it i)}  Parametrizations based on Gaussian wave 
functions with argument $\propto (1-x)\Delta^2$ \cite{Rad98}; 

\noindent
{\it ii)} Parametrizations based on Gaussian wave 
functions with argument $\propto (1-x)^n\Delta^2$, with $n \geq 2$,
 \cite{Bur_new};  

\noindent
{\it iii)} GPDs obtained using a diquark/spectator model consistent with a 
$1/k^4$ asymptotic dependence of the wavefunction.
\vspace{0.1cm}

We confirm the result initially obtained in Ref.\cite{Stoler} 
(Fig.\ref{fig5}) that $F_1$ at $\Delta > 2$ GeV can be reproduced 
only by allowing for a sufficiently large amount of large ${\bf k}$ 
(hard) components.
Cases {\it ii)}, and {\it iii)} satisfy this condition.  
Furthermore, we find that, for values of 
$\Delta^2 \approx 25-30$ GeV$^2$, the behavior predicted for the form factor 
(Figs.\ref{fig6}, \ref{fig9}, \ref{fig11}) is governed by:

\vspace{0.3cm}
\noindent 
{\it i)} Large $x$ ($x_{ave}(\Delta=5 \, {\rm GeV}) \approx 0.75$), 
$k \lesssim 0.7$ GeV in \cite{Rad98}; 

\noindent
{\it ii)} $x \rightarrow  0$, large $k$  in 
\cite{Bur_new}; 

\noindent 
{\it iii)} $x \approx 0.25$, $k \lesssim 2$ GeV in our diquark/spectator model.
\vspace{0.2cm}
 
The rms value of the proton radius, given by 
$\displaystyle \langle {\bf r}^2 \rangle^{1/2}$ in Eq.(\ref{r}), was 
calculated by a two-dimensional Fourier transform of the GPD, {\it i.e.} 
using the function $q(x,{\bf b})$ (Fig.\ref{fig7}). 
We pointed out that 
$\langle {\bf r} \rangle^{1/2} \rightarrow \langle {\bf y} \rangle^{1/2} $  for 
$x \rightarrow 1$, where 
${\bf y}$ is the variable defining the radius in Ref.\cite{Bur_new}, and 
$\langle {\bf r} \rangle^{1/2} \rightarrow \langle {\bf b} \rangle^{1/2} \neq \langle {\bf y} \rangle^{1/2}$ for $x\rightarrow 0$.
We unraveled  the following behavior (Fig.\ref{fig8}): 

\vspace{0.3cm}
\noindent
{\it i)} $\displaystyle \langle {\bf r^2} \rangle^{1/2}$ 
diverges for $x \rightarrow 0,1$ in \cite{Rad98}. 

\noindent
{\it ii)} $\displaystyle \langle {\bf r^2} \rangle^{1/2} \rightarrow 0$ 
for $x \rightarrow 1$ in \cite{Bur_new}.

\noindent
{\it iii)} $\displaystyle \langle {\bf r}^2 \rangle^{1/2}$ decreases from 
the value of $\approx 5$ GeV$^{-1}$ ($\approx 1$ fm), at 
$x \rightarrow 0$, to
$\approx 2.5$ GeV$^{-1}$ ($\approx 0.5$ fm), at $x \rightarrow 1$ 
in our model. 
\vspace{0.2cm}

From the behavior of the rms radius with $x$, one can surmise what components
in $b$ space the form factor is dominated by. 
While $\langle {\bf r}^2 \rangle ^{1/2}$ at $x>0.5$ 
is entirely given by $b \lesssim  2.5$ GeV$^{-1}$ ($0.5$ fm) 
for models characterized by a hard component ({\it ii)} and
{\it iii)}),
in case {\it i)} saturation is not reached (Fig.10), that is the rms radius
is large even at large $x$. 
On the other side, by studying the average value of $x$, $x_{ave}$, 
calculated using
the distribution $H$ vs. $\Delta$ (Figs.\ref{fig11} and \ref{fig12}), 
one finds that the hard type functions are not dominated by large
$x$ at large $\Delta$, rather all $x$ components seem to contribute
(Fig.\ref{fig12}). 

In summary, although it is becoming clear that the physics governing
the hadronic form factors, and more generally exclusive processes at 
large momentum transfer,  
is given by a non-trivial blend of  
soft and hard components, current descriptions using GPDs are not completely 
satisfactory.  

The next question is how one would be able to test the transverse spatial
structure of hadrons, given the fact that GPDs are quite elusive objects from
the experimental point of view. 
A potentially powerful means to 
single out the size of hadronic configurations
could be given by combining CT/nuclear filtering studies 
with GPDs as suggested in this paper. 
Based on the so far little explored 
concepts of IPPDFs and GPDs, 
we can now test the effect of nuclear filtering in an exclusive reaction
by building  a ``filter'' in the definition of the transparency, $T_A$. 
This procedure eliminates some of the model dependence implicit in many
predictions in a somewhat similar way to what proposed {\it e.g.} 
in Ref.\cite{JP_8}.
It however also provides a way of explicitly extracting the transverse
radial dependence of the hadron-hadron cross section in the nuclear medium.
While an analysis extended to all available data will be carried out in a 
upcoming paper \cite{LiuTan2}, we have presented here a few 
hypothetical scenarios
obtained with our procedure.   

In addition to the main results of this paper, interesting new 
developments will concern the extension of our calculations to 
the neutron and the polarized case, as well as the 
exploration of the skewedness, $\xi$, dependence.
Most importantly, the use of nuclei and CT experiments might unravel an 
alternative method to measure GPDs by evaluating nuclear dependent 
contributions beyond the standard assumption of factorization 
into a nucleon times a nuclear part for
the nuclear scattering amplitude.
Such terms can be shown to be directly proportional to GPDs and 
a study examining their relative importance as $A$-dependent 
contributions is on its way \cite{LiuTan2}.
Both the studies initiated in this paper and the suggested future 
studies could be tested ideally both at Jefferson Lab \cite{Jlab_01_107} 
including its 12 GeV upgrade, and 
at newly planned  
Electron Ion Colliders with high luminosity and energy \cite{EIC}.  

In conclusion, studies like the one presented
here using the new concept of GPDs, will both improve our knowledge of 
nuclear filtering phenomena  
and allow for a more detailed understanding of the transverse 
components involved at large momentum transfer.  
In particular, we hope to have provided a connection between
${\bf b}-$ and ${\bf k}-$ space that will help 
to systematically address both the role 
of Sudakov effects in
the endpoints of the hadronic wave function \cite{Kundu,Hoyer}, 
and the role of power corrections in the large longitudinal
momentum regions hinted in \cite{Bur_new}. 

\acknowledgments
We thank Karo Oganessyan for comments on the manuscript. 
This work is supported by the U.S. Department
of Energy grant no. DE-FG02-01ER41200. 

\appendix
\section{}

The coefficients defining $H(x,{\bf \Delta})$ in Eq.(\ref{H_par}) 
are given by:
\begin{subequations}
\begin{eqnarray}
p1 & = & 2.780 \, x^{0.2} (1-x)^{4.2}  (1+0.5 \, x^{0.3}) \\
p2 & = & \left \{ \begin{array}{c}
 0.5695 -0.1896 \, x + 0.6885 \, x^2  \; \; x \leq 0.4 \\
0.605 \; \;  x>0.4  \end{array} \right.
\\
p3 & = & 1.118 \, x^{-0.098} (1-x)^{2.24}  \\
p4 & = & \left \{ \begin{array}{c} 
0  \; \; x \leq 0.87 \\
-0.1076 x + 0.1079 \; \; x>0.87 \end{array} \right. 
\end{eqnarray}
\end{subequations}




\vspace{5cm} 
\newpage
\begin{figure}
\includegraphics[width=16.cm]{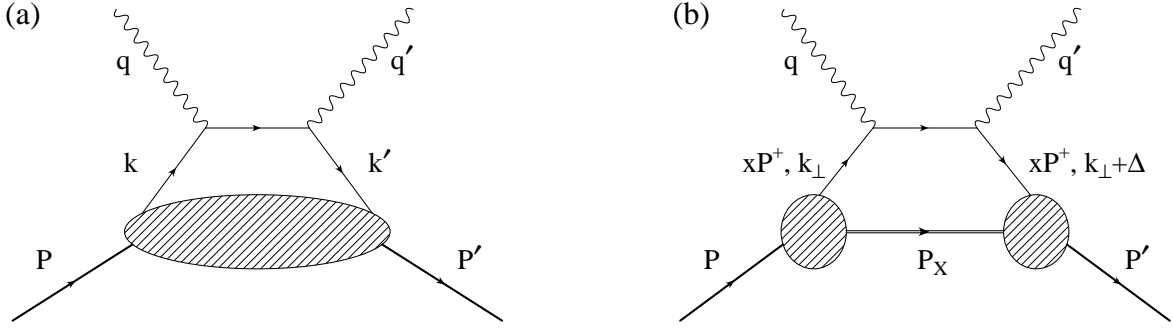}
\caption{(a) Amplitude for DVCS at leading order in $Q^2$; 
(b) The same amplitude in a two component model.} 
\label{fig1}
\end{figure}
\begin{figure}
\includegraphics[width=15.cm]{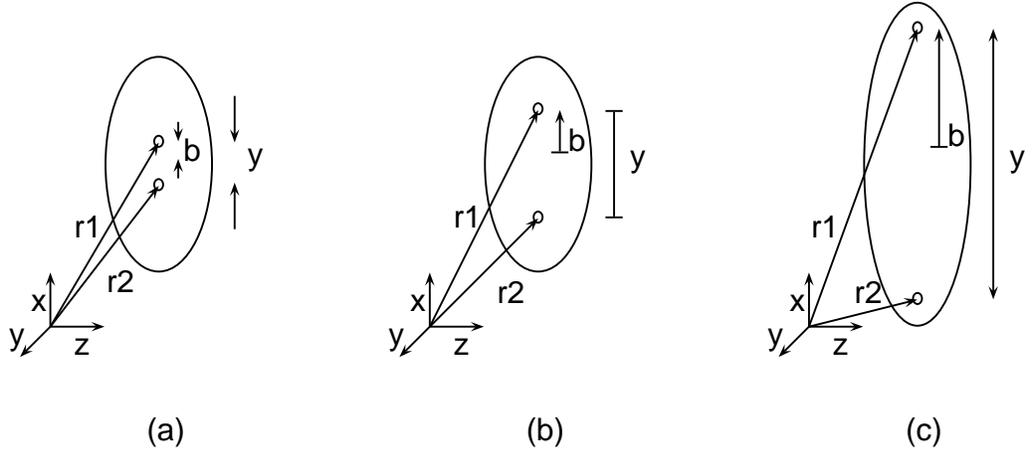}
\vspace{-3cm}
\caption{The transverse distance between the active quark (upper blob) and
the center of momentum of the spectator quarks (lower blob) -- ${\bf y}$;  
between the active quark and the center of momentum of the hadron 
(full circle) -- 
${\bf b}$, and between the spectator quarks and the center of mass of the
hadron -- ${\bf b} x/(1-x) \equiv {\bf y}-{\bf b}$.
$r_1$ and $r_2$ are the space vectors for the active quark and the 
center of momentum of the spectator quarks, respectively, taken in the 
system of coordinates $(x,y,z)$. For simplicity, ${\bf b}$ is drawn parallel 
to the $x$-axis.  
Assuming IPPDFs that yield values of $\langle {\bf b}^2 \rangle$ 
decreasing as $x \rightarrow 1$, there are three possible behaviors for ${\bf y}$
and ${\bf r}$ (Eq.(\ref{r})):
{\bf (a)} ${\bf b} \propto (1-x)^\gamma,\gamma>{1}$
$ \Rightarrow {\bf y}, {\bf r}$ decrease as $x$ increases;
{\bf (b)} ${\bf b} \propto(1-x)^\gamma,\gamma={1}$ $ \Rightarrow {\bf y}, {\bf r}$ 
are constants independent of $x$;
{\bf (c)} ${\bf b} \propto (1-x)^\gamma,\gamma<{1}$ 
$ \Rightarrow {\bf y}, {\bf r}$ increase as $x$ increases.}
\label{fig2}
\end{figure}

\begin{figure}
\includegraphics[width=16.cm]{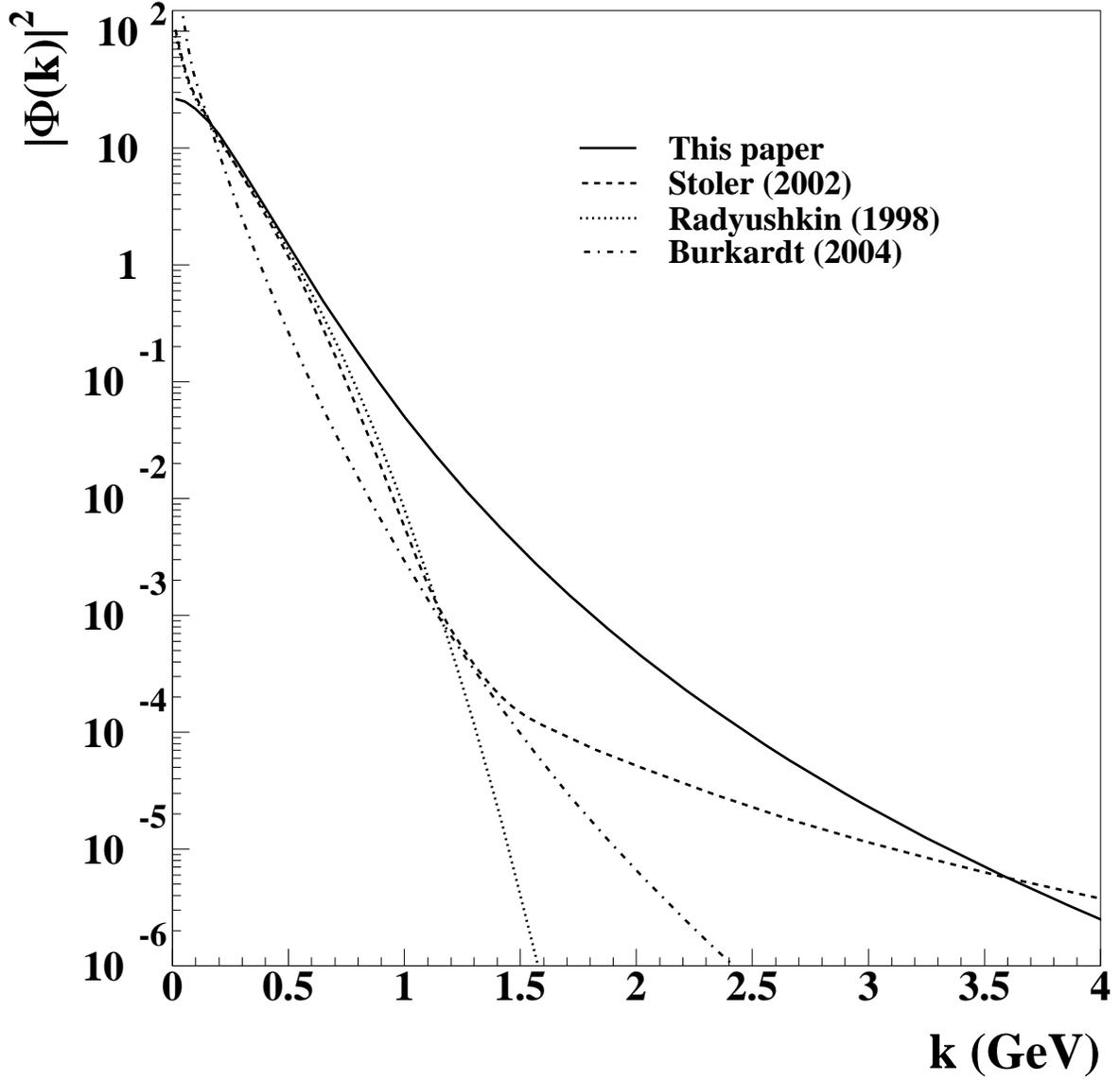}
\caption{The function 
$\displaystyle \mid \Phi({\bf k}) \mid^2$, defined
in Eq.(\ref{funck}). Full line: model calculation discussed
in this paper; dotted line: the soft wave function introduced in
Ref.\cite{Rad98}; dashed line: model of Ref.\cite{Stoler}, including
a hard component in $\Phi$; dot-dashed line, the ``semi-hard'' distribution
proposed in Ref.\cite{Bur_new}.}
\label{fig3}
\end{figure}

\begin{figure}
\includegraphics[width=16.cm]{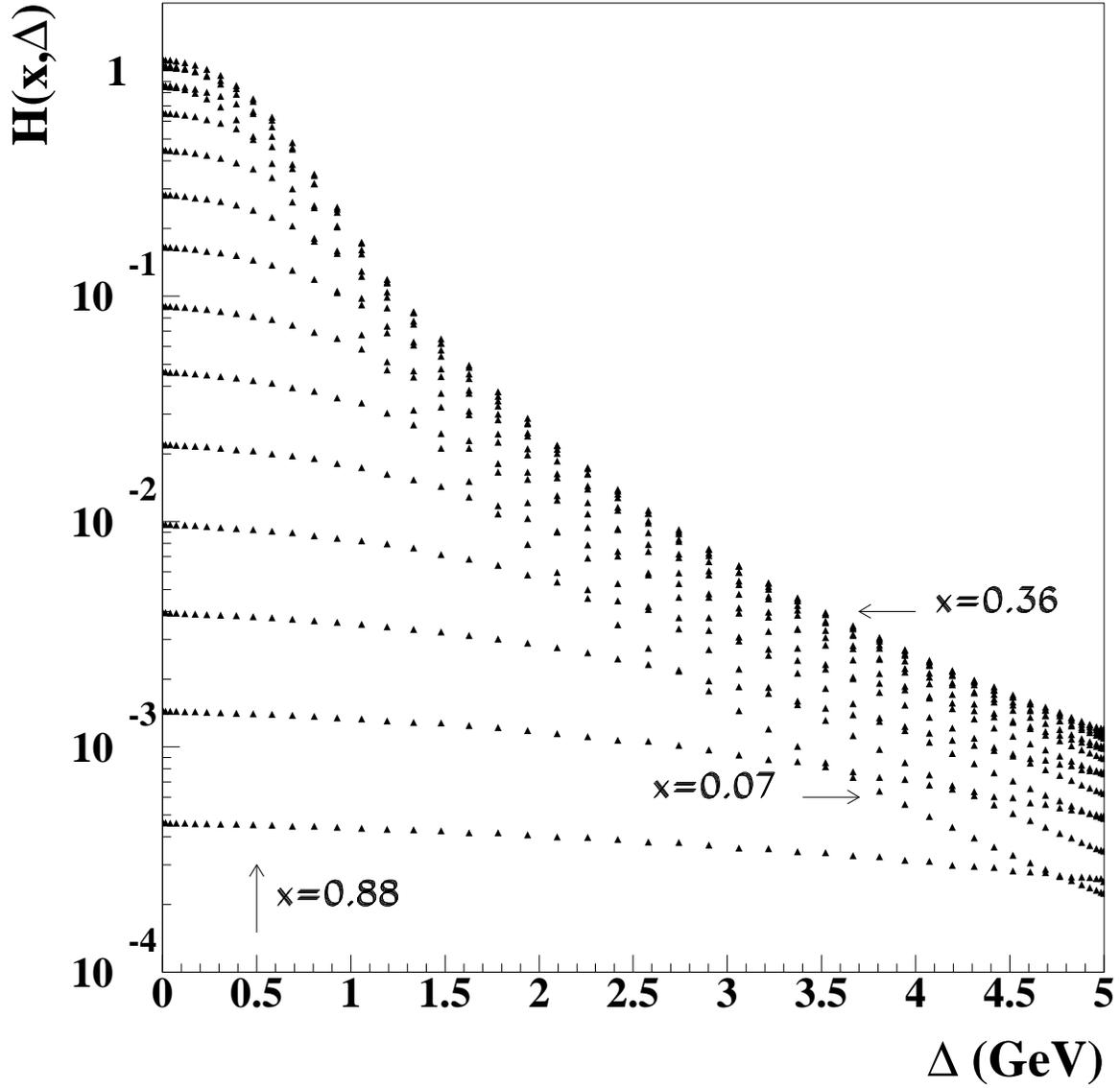}
\caption{The GPD, $H(x,{\bf \Delta})$, calculated according to
Eqs.(\ref{gpd3}) and (\ref{H_par}), plotted vs. $\Delta$ for
different values of $x$. Typical values, $x=0.07, 0.36, 0.88$, are shown 
by the arrows.}

\label{fig4}
\end{figure}

\begin{figure}
\includegraphics[width=16.cm]{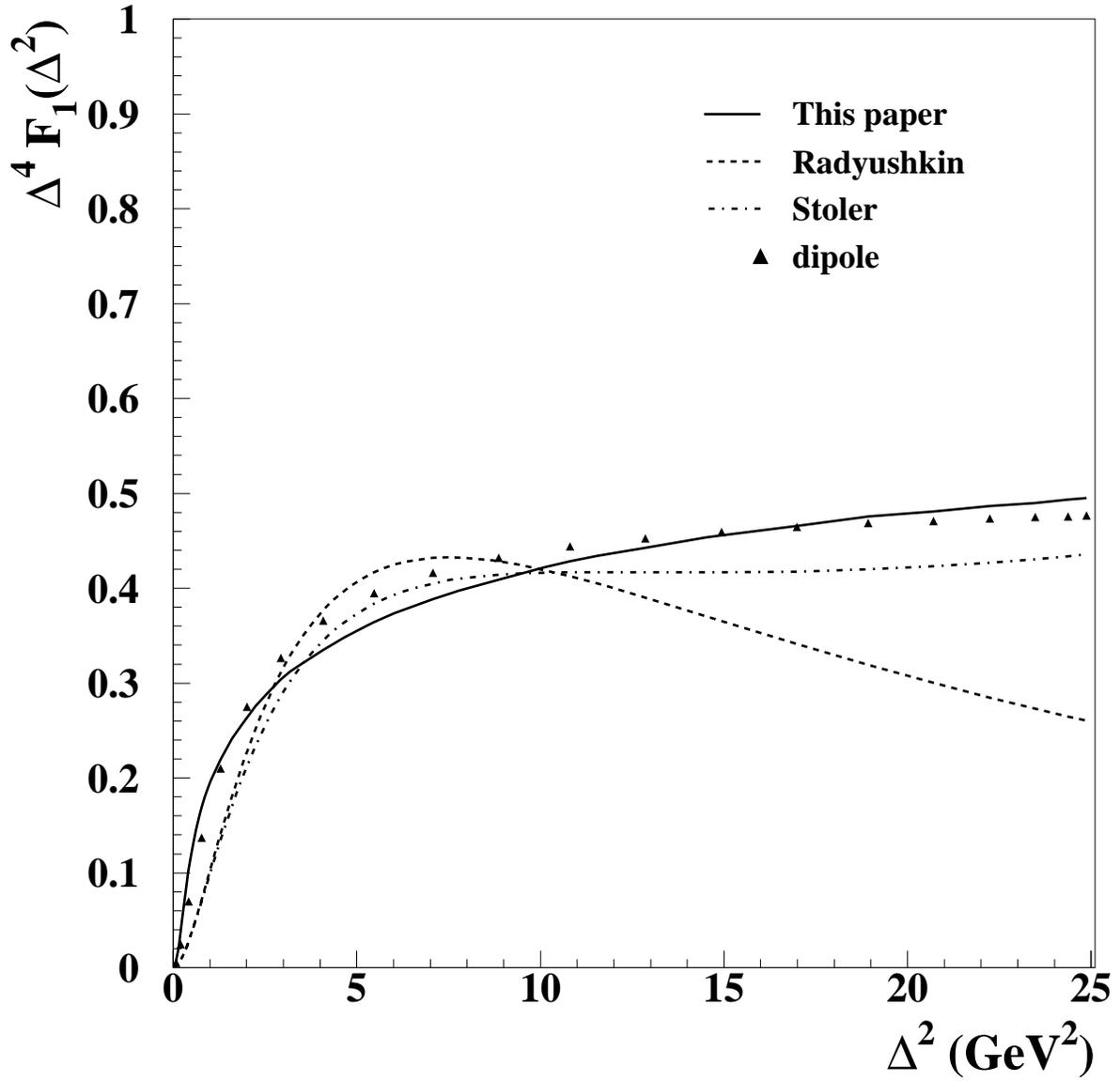}
\caption{The form factor, $F_1$ multiplied by $\Delta^4$, 
calculated in three different
models: full line, this paper; dashed line, \cite{Rad98}; dot-dashed
line, \cite{Stoler}. Results are compared with the dipole form factor
(triangles).}

\label{fig5}
\end{figure}

\begin{figure}
\includegraphics[width=16.cm]{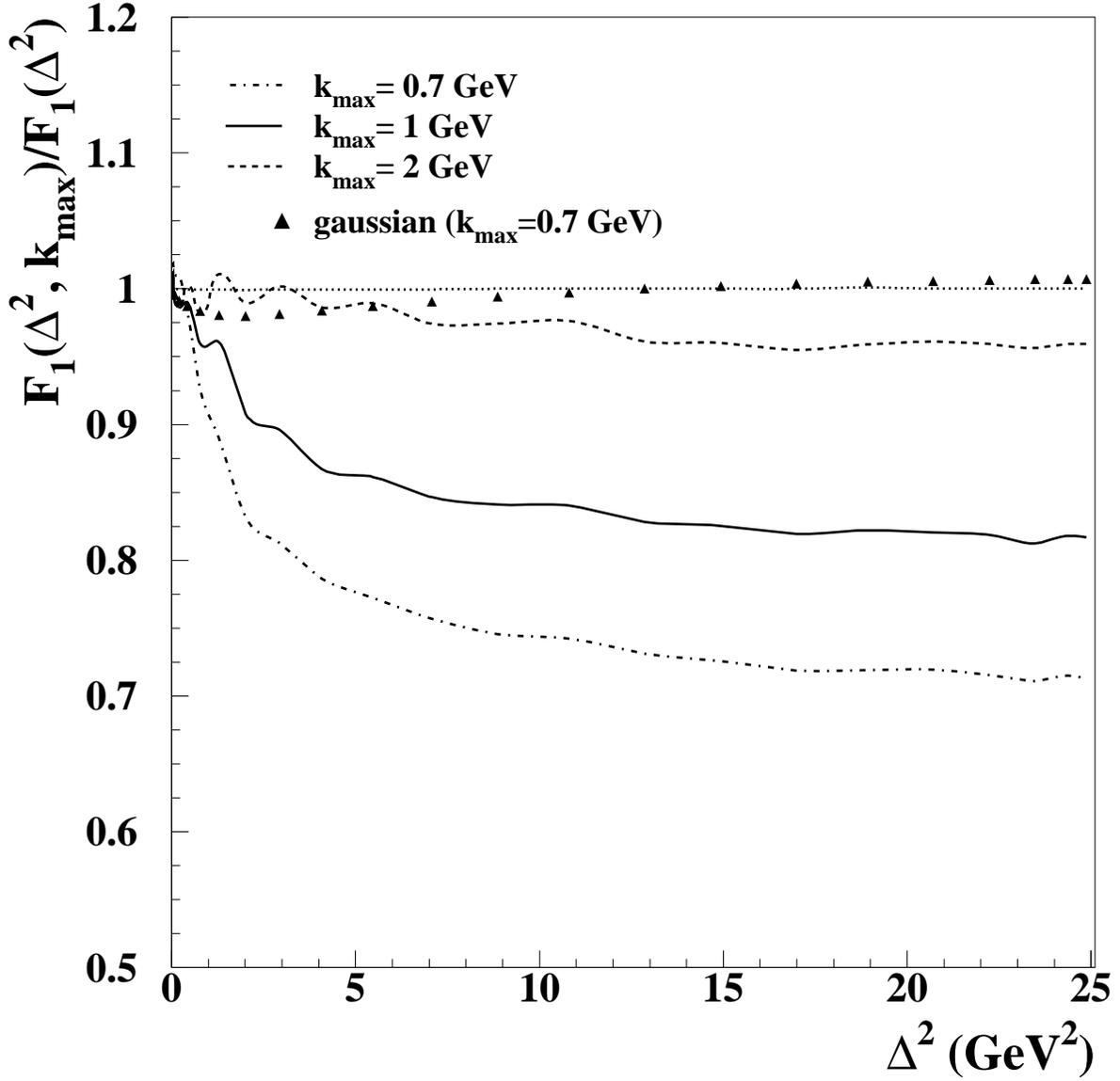}
\caption{Dominance of different $k$ components in $F_1(\Delta^2)$, showed 
using the ratio $F_1(\Delta^2, k_{max})/F_1(\Delta^2)$. The numerator 
was obtained by setting the upper limit of 
integration in Eq.(\ref{ff3}) to different values of $k \equiv k_{max}$. 
The dot-dashed line is obtained for $k_{max}=0.7$ GeV; the full line 
corresponds to $k_{max}=1 $ GeV, and the dashed line to $k_{max}=2$ GeV. 
The triangles, obtained by setting $k_{max}=0.7$ in the model of 
Ref.\cite{Rad98}, clearly show the saturation of the ratio at low values
of $k$.}

\label{fig6}
\end{figure}

\begin{figure}
\includegraphics[width=16.cm]{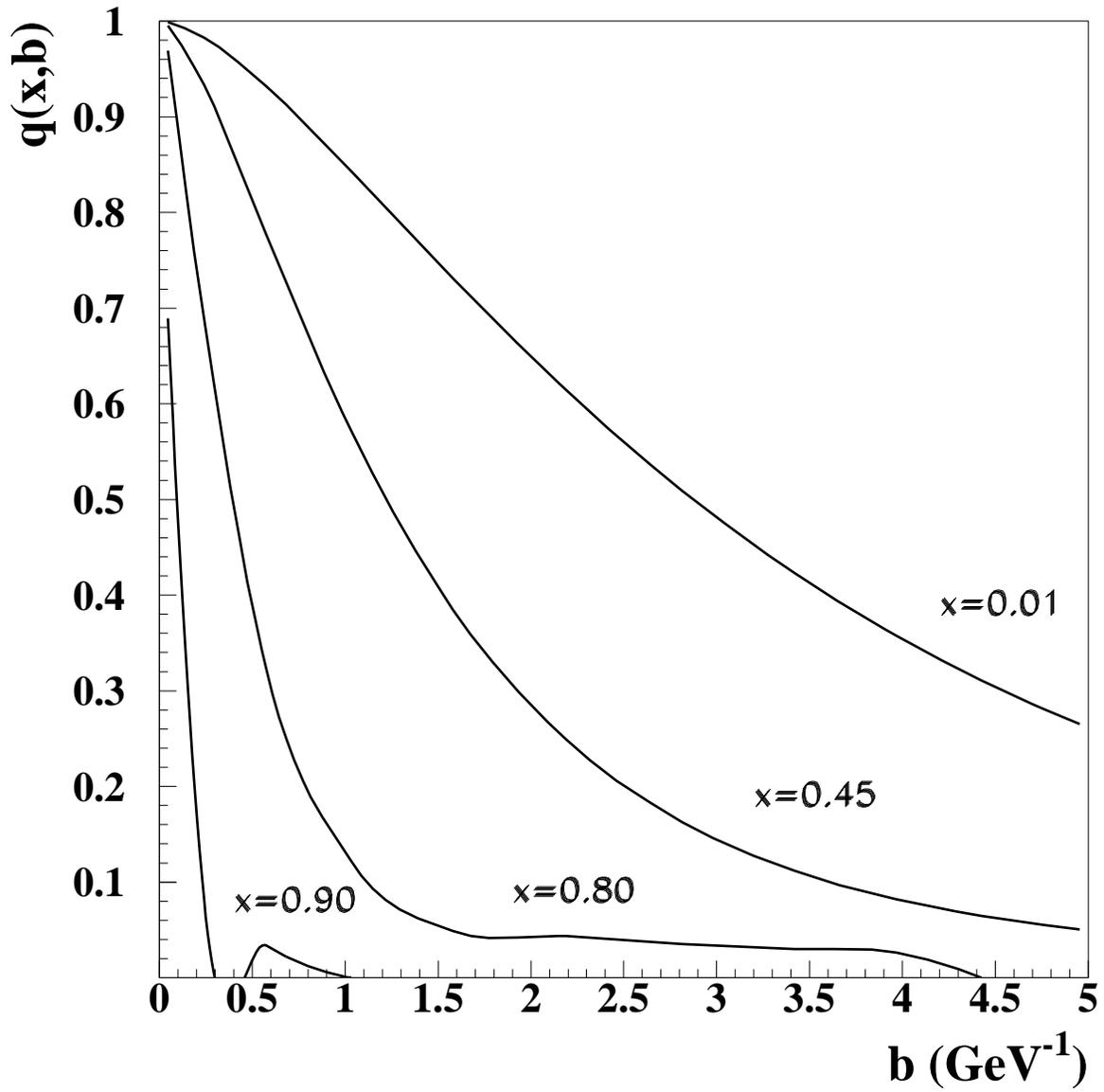}
\caption{IPPDF, $q(x,b)$ obtained by two-dimensional Fourier 
transforming Eq.(\ref{H_par}), plotted vs. $b$. 
The curves shown in the figure correspond to different values of $x:
0.01, 0.45, 0.8, 0.9$.}

\label{fig7}
\end{figure}

\begin{figure}
\includegraphics[width=9.cm]{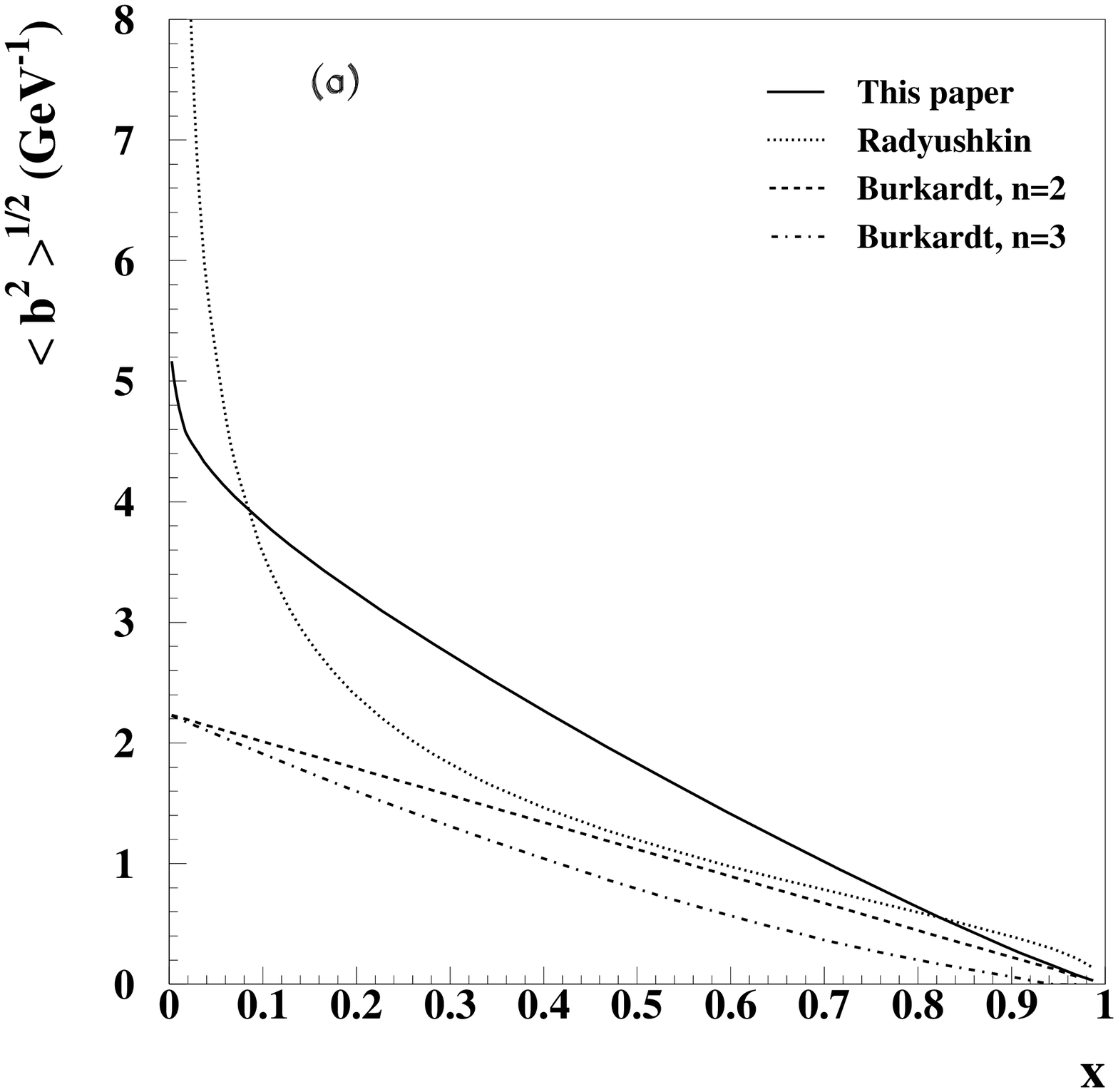}
\includegraphics[width=9.cm]{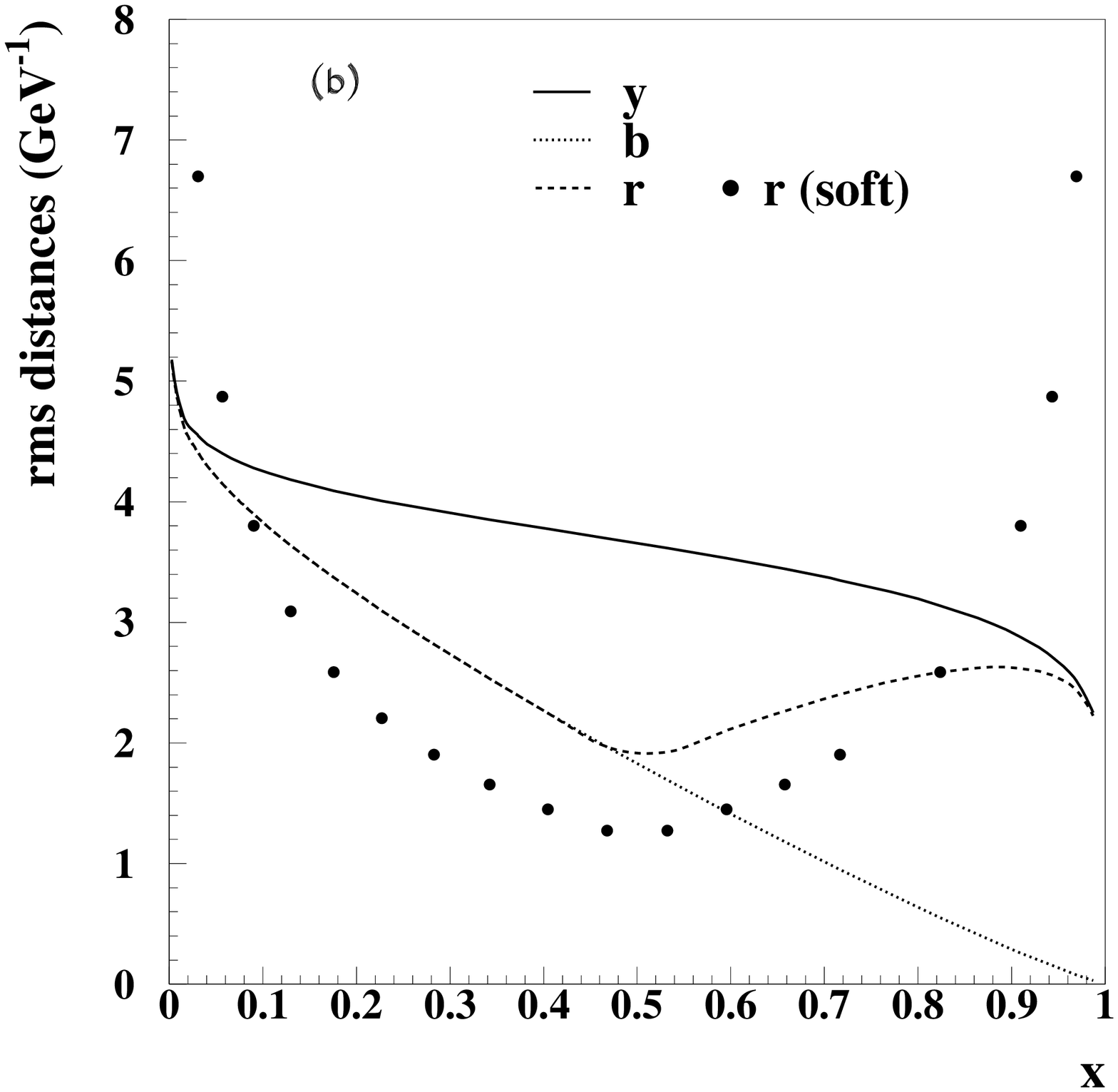}
\caption{Modeling the proton radius. 
{\bf (a)} $\langle b^2 \rangle^{1/2}$ in GeV$^{-1}$, Eq.(\ref{b}), plotted vs. $x$, 
calculated in different models: full line, this paper; dotted line \cite{Rad98}; the 
dashed and dot-dashed lines were calculated
using the parametrization proposed in \cite{Bur_new}. This parametrization
describes the $\Delta^2$ behavior using a gaussian as in \cite{Rad98}, 
however the argument which is $\propto (1-x) \Delta^2$ in \cite{Rad98}
is suggested to be $(1-x)^n \Delta^2$, with $n \geq 2$ in \cite{Bur_new}.
The dashed line and the dot-dashed line were obtained by setting $n=2$ and 
$n=3$, respectively.
{\bf (b)} Transverse radial quantities: full line, $y \equiv \langle {\bf y}^2 \rangle^{1/2}$, 
Eq.(\ref{y}); dots, $b \equiv \langle {\bf b}^2 \rangle^{1/2}$, Eq.(\ref{b}), 
dashes, $r \equiv \langle {\bf r}^2 \rangle^{1/2}$, Eq.(\ref{r}). 
All quantities are calculated in the diquark model. For comparison
$r$ calculated in the model of \cite{Rad98}, based on a ``soft'' type wave function, is also
shown.}

\label{fig8}
\end{figure}

\begin{figure}
\includegraphics[width=9.cm]{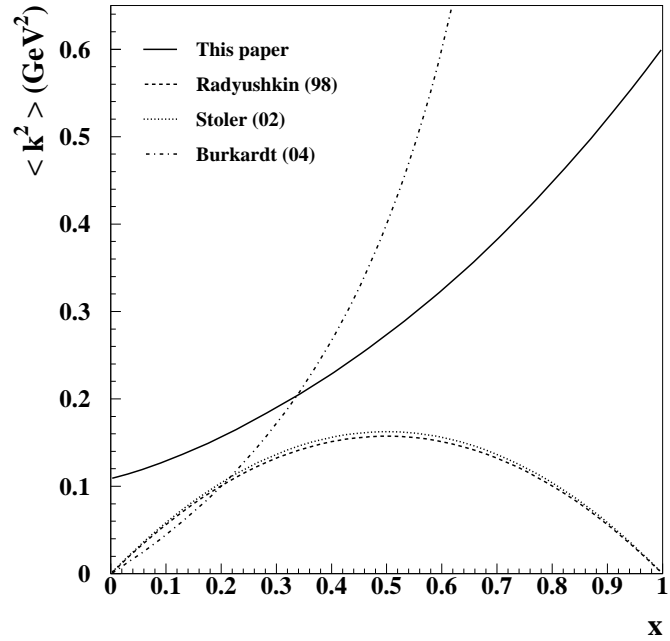}
\caption{Average intrinsic transverse momentum squared, Eq.(\ref{k}), 
plotted as a function of $x$ using different wave functions.
Full line: model calculation discussed
in this paper; dotted line: 
Ref.\cite{Rad98}; dashed line: Ref.\cite{Stoler}; dot-dashed line, 
the ``semi-hard'' distribution proposed in Ref.\cite{Bur_new}.}

\label{fig9}
\end{figure}

\begin{figure}
\includegraphics[width=9.cm]{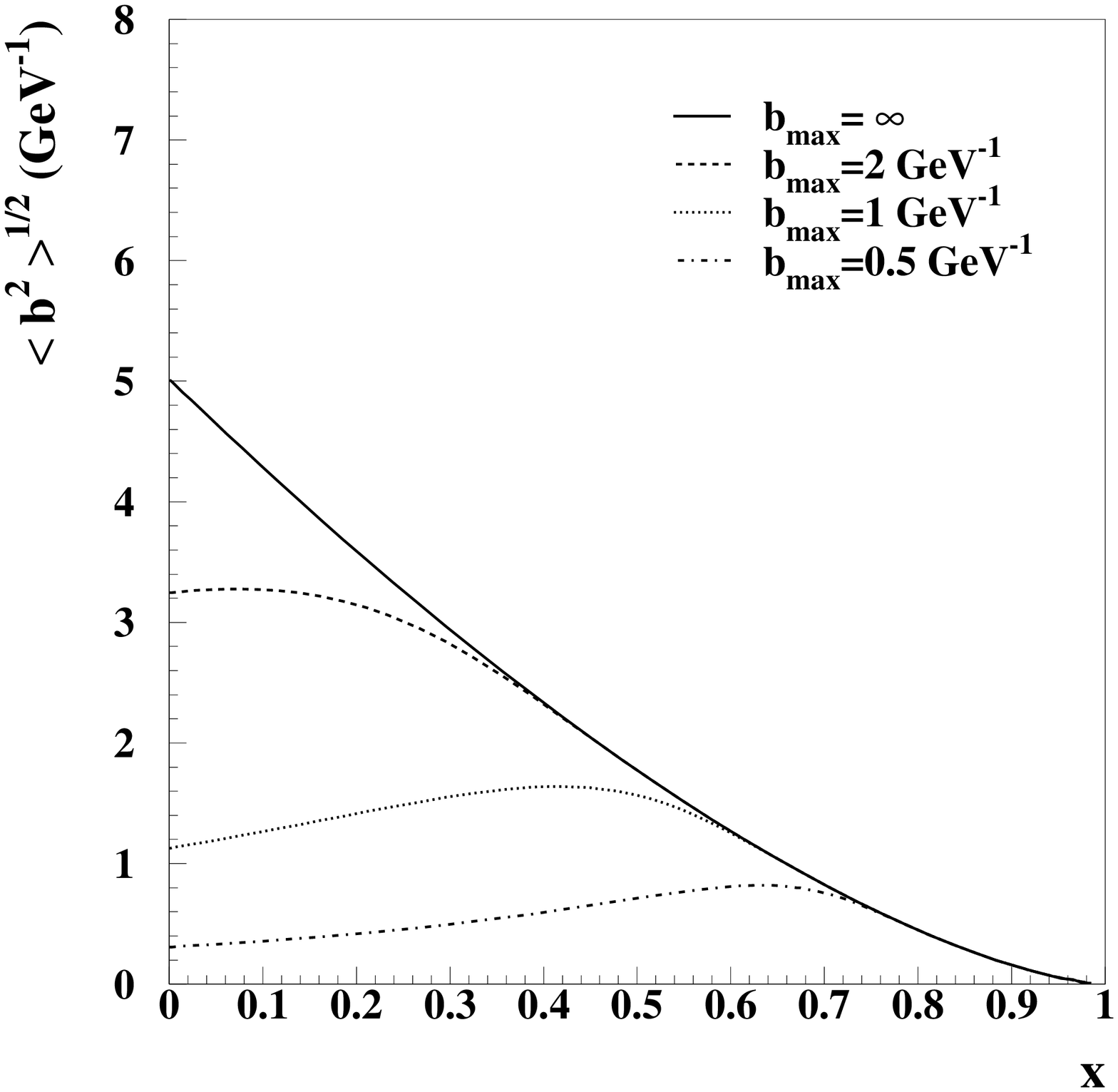}
\includegraphics[width=9.cm]{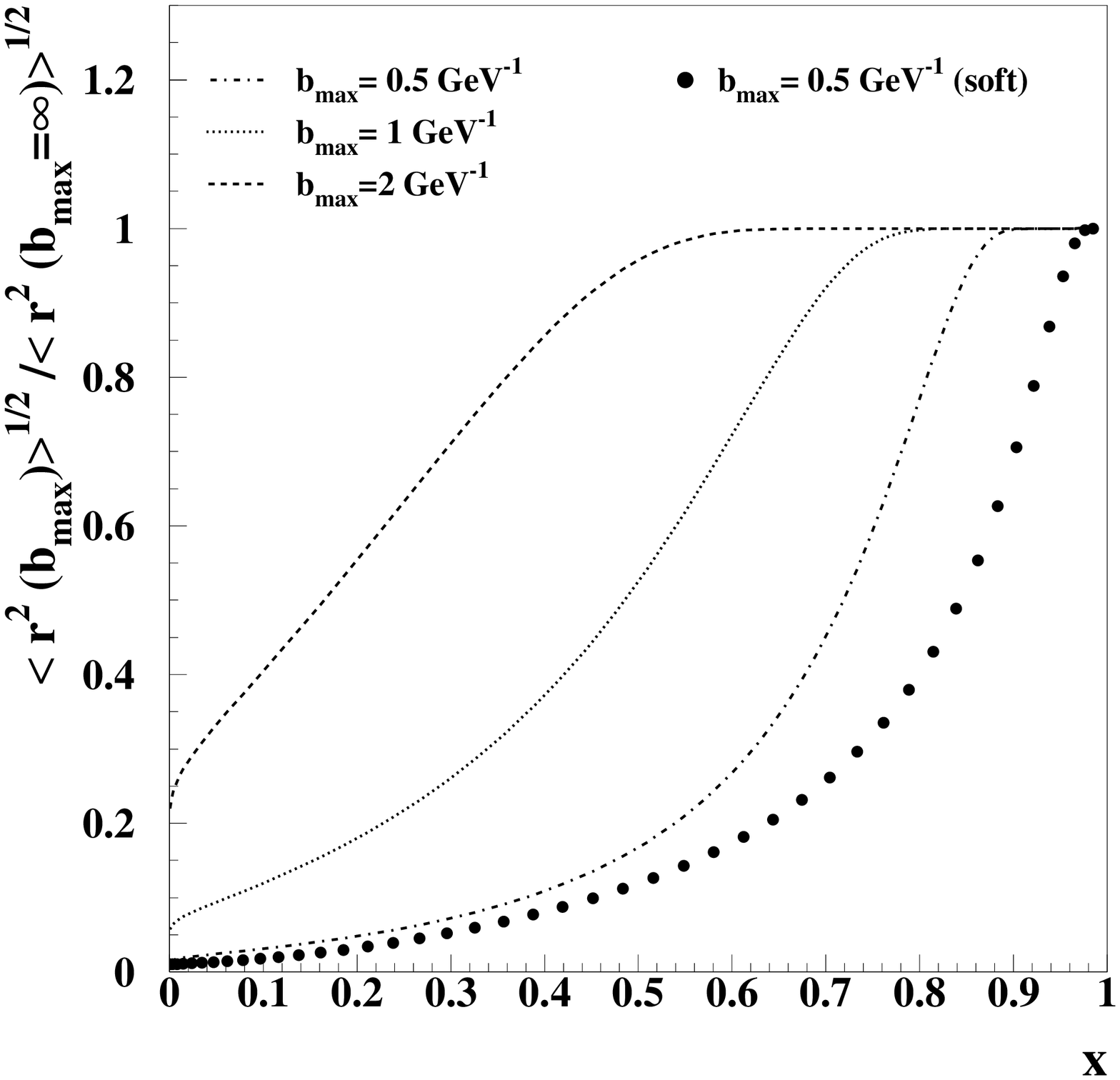}
\caption{Contribution of different transverse distances to the 
rms impact parameter $\langle b^2 \rangle^{1/2}$, Eq.(\ref{b}).
Upper panel: Different contributions are calculated with the spectator/diquark model. They are 
evaluated  setting the upper limit of 
integration in Eq.(\ref{b}) to different values of $b \equiv b_{max}$. The full line 
is obtained integrating over the whole range; the dashed line corresponds to 
$b_{max} = 2$ GeV$^{-1}$;  the dotted line to 
$b_{max} = 1$ GeV$^{-1}$; the dot-dashed line to 
$b_{max} = 0.5$ GeV$^{-1}$. Lower panel: ratio of the partial contributions displayed 
in the upper panel to the calculation using the full range of $b$. The continuous lines
were obtained using the function in the upper panel. The big dots were obtained using 
the model of Ref.\cite{Rad98}. Notice that the ratio saturates much faster as a function
of $x$ for the ``hard'' distribution adopted in this paper, than for the model of 
\cite{Rad98}. }

\label{fig10}
\end{figure}
\begin{figure}
\includegraphics[width=12.cm]{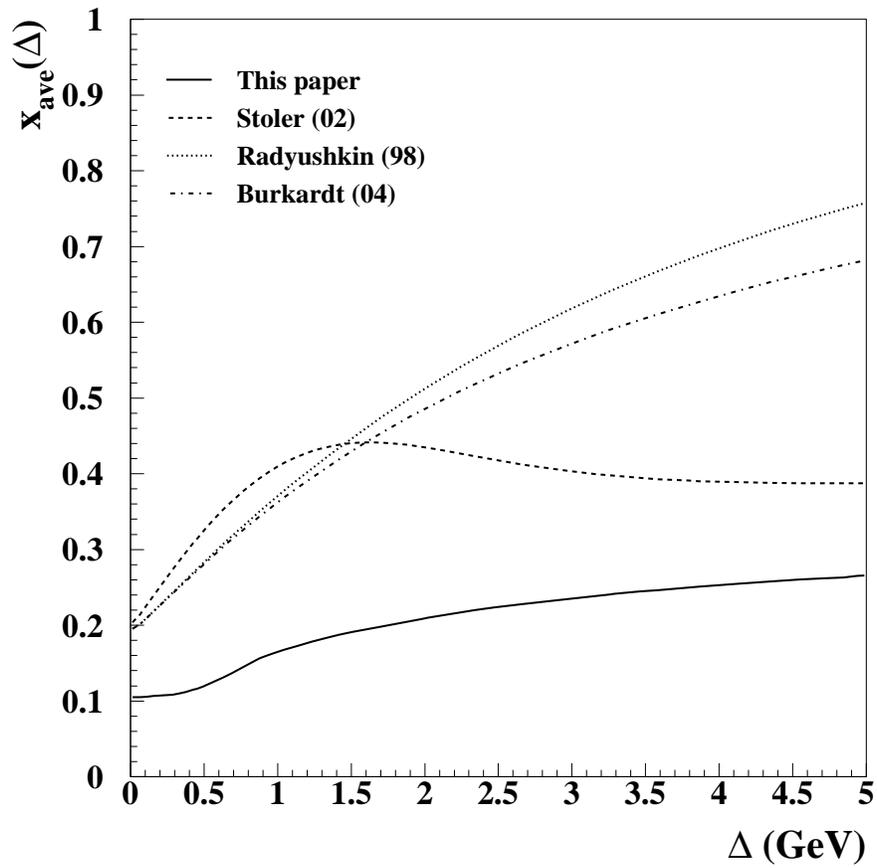}
\caption{Average value of $x$, $x_{ave}$, calculated using different wave functions in $H(x,\Delta)$ 
as a function of $\Delta$. The full line corresponds to the spectator model; the dashed line
to the model of Ref.\cite{Stoler}, the dotted line to Ref.\cite{Rad98}, the dot-dashed line to 
Ref.\cite{Bur_new}, for $n=2$ (see also Fig.\protect\ref{fig8}). }

\label{fig11}
\end{figure}

\begin{figure}
\includegraphics[width=16.cm]{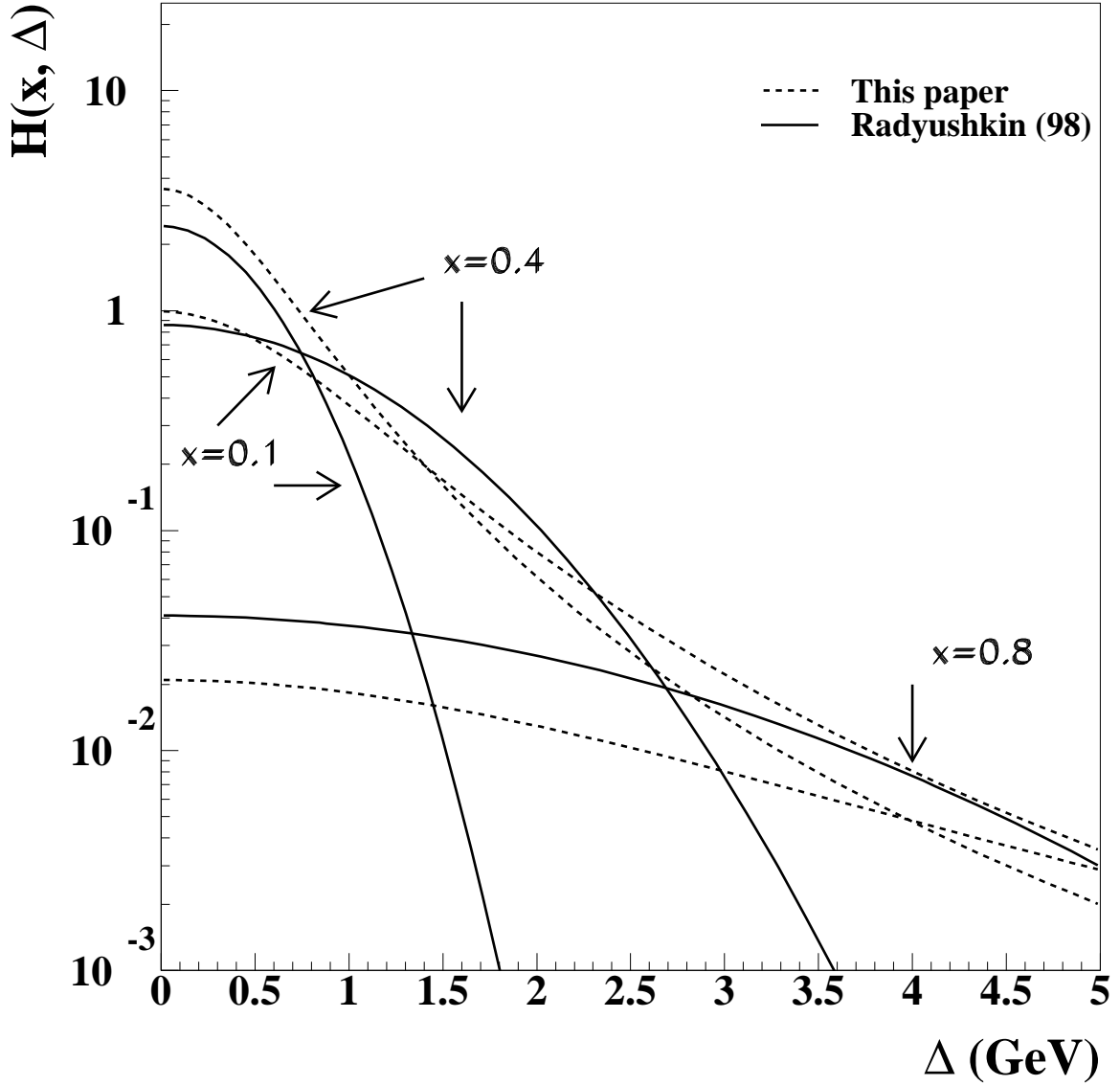}
\caption{Comparison of different $x$ components contributions
to $H(x, \Delta)$ in the spectator model (full lines), and in the model of Ref.\cite{Rad98}
(dashed lines).
The latter, although being soft, shows the dominance of (small) large $x$ components at 
(small) large $\Delta$. In the spectator model, regardless of its hard $k$ components, 
also small values of $x$ contribute to large $\Delta$. The arrows in the figure indicate
the three different values of $x$ for which the curves were calculated. }

\label{fig12}
\end{figure}

\begin{figure}
\includegraphics[width=12.cm]{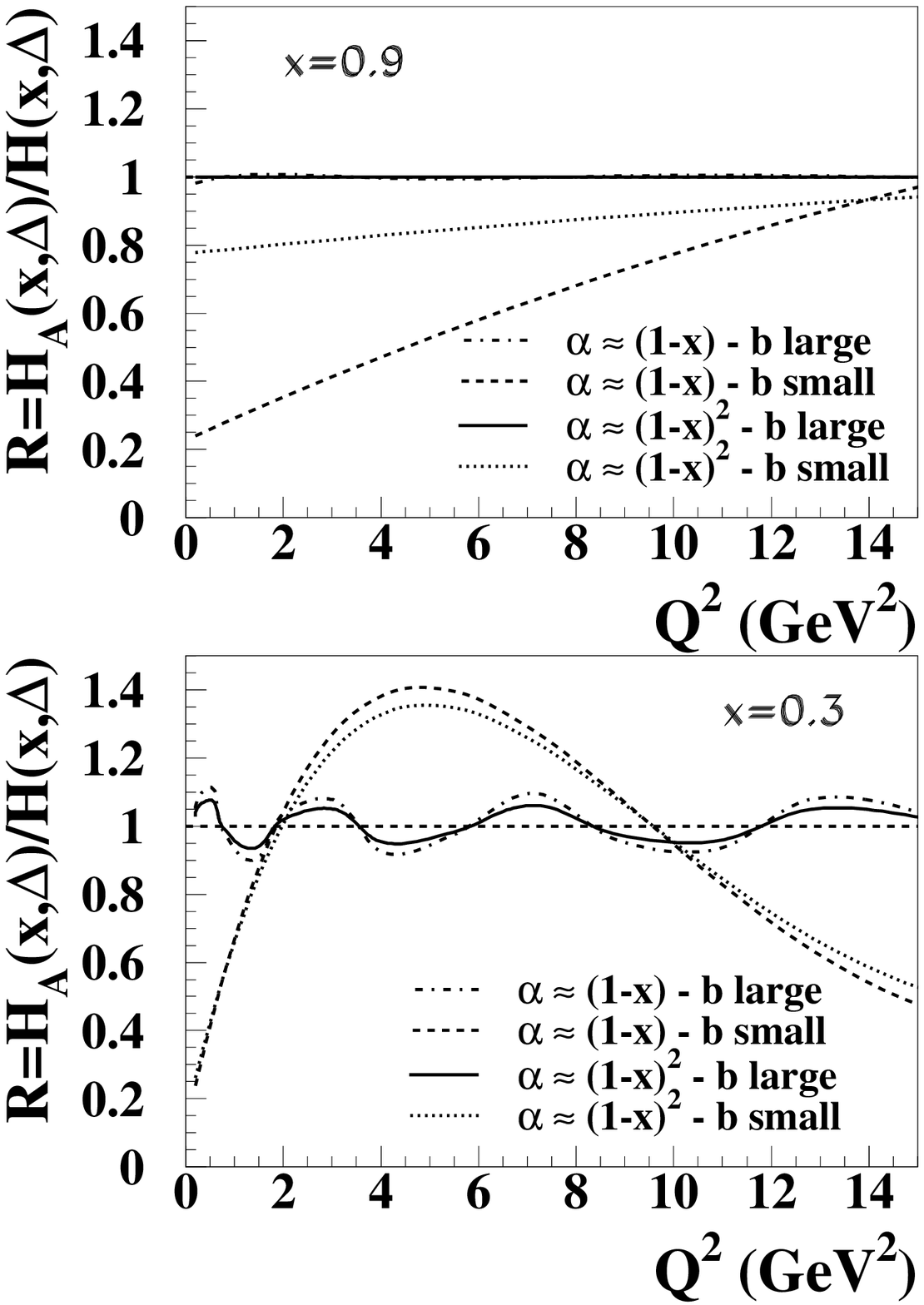}
\caption{Possible scenarios for the effect of nuclear filtering and 
the onset of CT based on a simplified
model described in the text.
The ratio $R$ of GPDs in a nucleus and in free space is plotted vs.
$Q^2 \equiv \Delta^2$, for two different values of $x$: $x=0.9$, upper 
panel and $x=0.3$, lower panel.
For each panel we consider: {\bf (a)} $\alpha \propto (1-x)/x$, as 
in Ref.\cite{Rad98} (dot-dashed and dashed lines); {\bf (b)} 
$\alpha \propto (1-x)^2$, as in Ref.\cite{Bur_new} (full and dotted
lines). Moreover, the full and dot-dashed line correspond
to a large value of the filter's size, $b_{max}$, the dashed and dotted 
lines to a small value of $b_{max}$. The effect of the latter is to
produce a suppression in the ratio, which is more pronounced at large
$x$ for distributions with a large radius \cite{Rad98}. }  
\label{fig13}
\end{figure}


\begin{thebibliography}{99}


\bibitem{BroMul}S.~J.~Brodsky and A.~H.~Mueller,
Phys.\ Lett.\ B {\bf 206}, 685 (1988).

\bibitem{Bro_rev} S.~J.~Brodsky,
arXiv:hep-ph/0311355, {\it and references therein}.

\bibitem{JP_01} P.~Jain and J.~P.~Ralston,
Pramana {\bf 57}, 433 (2001)
[arXiv:hep-ph/0103131].

\bibitem{RalPir} J.~P.~Ralston and B.~Pire,
Phys.\ Rev.\ Lett.\  {\bf 65}, 2343 (1990).

\bibitem{Kundu} B.~Kundu, J.~Samuelsson, P.~Jain and J.~P.~Ralston,
Phys.\ Rev.\ D {\bf 62}, 113009 (2000); {\it ibid} 
AIP Conf.\ Proc.\  {\bf 549}, 455 (2002)
[arXiv:hep-ph/0008194].

\bibitem{exp_CT} A.~S.~Carroll {\it et al.},
Phys.\ Rev.\ Lett.\  {\bf 61}, 1698 (1988); 
N.~Makins {\it et al.},
Phys.\ Rev.\ Lett.\  {\bf 72}, 1986 (1994); T.~G.~O'Neill {\it et al.},
Phys.\ Lett.\ B {\bf 351}, 87 (1995); K.~Garrow {\it et al.},
Phys.\ Rev.\ C {\bf 66}, 044613 (2002).

\bibitem{JaiPirRal} P.~Jain, B.~Pire and J.~P.~Ralston,
Phys.\ Rept.\  {\bf 271}, 67 (1996)

\bibitem{FMS} L.L. Frankfurt, G.A. Miller, M.I. Strikman, Ann. Rev. Nucl. 
Part. Sci. {\bf 44}, 501 (1994); {\it ibid}, Comm. Nucl. Part. Phys.,
{\bf 21} 1, (1992).

\bibitem{NNN} N.N. Nikolaev, Comm. Nucl. Part. Phys.,
{\bf 21}, 41 (1992).

\bibitem{JP_7} P.~Jain, B.~Kundu and J.~P.~Ralston,
Phys.\ Rev.\ D {\bf 65}, 094027 (2002).

\bibitem{Boris} B.Z. Kopeliovich {\it et al.}, Phys. Rev. {\bf C65},
035201 (2002). 

\bibitem{HERMES_CT} A.~B.~Borissov  [HERMES Collaboration],
Nucl.\ Phys.\ A {\bf 711}, 269 (2002).
 
\bibitem{Gao} L.~Y.~Zhu {\it et al.}  [Jefferson Lab Hall A Collaboration],
Phys.\ Rev.\ Lett.\  {\bf 91}, 022003 (2003).

\bibitem{Dutta} D.~Dutta {\it et al.}  [Jefferson Lab E940104 Collaboration],
Phys.\ Rev.\ C {\bf 68}, 021001 (2003).

\bibitem{Ait} E.M. Aitala {\it et al.}, Phys.\ Rev.\ Lett.\  {\bf 86}, 4773 (2001).

\bibitem{Hoyer} P.~Hoyer, J.~T.~Lenaghan, K.~Tuominen and C.~Vogt,
arXiv:hep-ph/0210124.

\bibitem{Burka} M.~Burkardt,
Int.\ J.\ Mod.\ Phys.\ A {\bf 18}, 173 (2003)
[arXiv:hep-ph/0207047].


\bibitem{Diehl_03} M.~Diehl,
Eur.\ Phys.\ J.\ C {\bf 25}, 223 (2002)
[Erratum-ibid.\ C {\bf 31}, 277 (2003)]

\bibitem{BelJiYuan} A.~V.~Belitsky, X.~d.~Ji and F.~Yuan,
arXiv:hep-ph/0307383.

\bibitem{Ji} X.~D.~Ji,
Phys.\ Rev.\ Lett.\  {\bf 78}, 610 (1997)

\bibitem{Rad} A.~V.~Radyushkin,
Phys.\ Lett.\ B {\bf 380}, 417 (1996).

\bibitem{Goe} K.~Goeke, M~.~V.~Polyakov and M.~Vanderhaeghen,
Prog.\ Part.\ Nucl.\ Phys.\  {\bf 47}, 401 (2001)

\bibitem{Die} M.~Diehl,
Phys.\ Rept.\  {\bf 388}, 41 (2003).

\bibitem{LiuTan2} S.~Liuti and S.~K.~Taneja, {\it in preparation}. 

\bibitem{Soper} D.~E.~Soper,
Phys.\ Rev.\ D {\bf 15}, 1141 (1977).

\bibitem{BHPS} S.~J.~Brodsky, P.~Hoyer, N.~Marchal, S.~Peigne and F.~Sannino,
Phys.\ Rev.\ D {\bf 65}, 114025 (2002).

\bibitem{femto} J.~P.~Ralston and B.~Pire,
Phys.\ Rev.\ D {\bf 66}, 111501 (2002).

\bibitem{Lep} G.~P.~Lepage, S.~J.~Brodsky, T.~Huang and P.~B.~Mackenzie,
CLNS-82/522,
{\it Invited talk given at Banff Summer Inst. on Particle Physics, Banff, Alberta, Canada, Aug 16-28, 1981}.

\bibitem{GroLiu} F.~Gross and S.~Liuti, 
Phys.\ Rev.\ C {\bf 45}, 1374 (1992).

\bibitem{Mul} R.~Jakob, P.~J.~Mulders and J.~Rodrigues,
Nucl.\ Phys.\ A {\bf 626}, 937 (1997), {\it and references therein}.

\bibitem{KPW} S.~A.~Kulagin, G.~Piller and W.~Weise,
Phys.\ Rev.\ C {\bf 50}, 1154 (1994).



\bibitem{Kroll} P.~Kroll, 
Proceedings of the workshop on 
{\it ``Exclusive processes at high momentum transfer}, Editors
A. Radyushkin and P. Stoler, World Scientific (2002), pp.214-224;
arXiv:hep-ph/0207118, and {\it references therein}.

\bibitem{Lon} A.~Gardestig, A.~P.~Szczepaniak and J.~T.~Londergan,
Phys.\ Rev.\ D {\bf 68}, 034005 (2003).

\bibitem{GRV_OLD} M.~Gluck and E.~Reya,
Nucl.\ Phys.\ B {\bf 130}, 76 (1977).

\bibitem{Rad98} A.~V.~Radyushkin, Phys.\ Rev.\ D {\bf 58}, 114008 (1998).

\bibitem{Stoler} P. Stoler, Phys.\ Rev.\ D {\bf 65}, 053013 (2002).

\bibitem{Bur_new} M.~Burkardt, arXiv:hep-ph/0401159.

\bibitem{dvcs_HERMES} A.~Airapetian {\it et al.}  [HERMES Collaboration],
Phys.\ Rev.\ Lett.\  {\bf 87}, 182001 (2001).

\bibitem{dvcs_Jlab} S.~Stepanyan {\it et al.}  [CLAS Collaboration],
Phys.\ Rev.\ Lett.\  {\bf 87}, 182002 (2001)

\bibitem{Nus} S.~Nussinov,
Phys.\ Rev.\ Lett.\  {\bf 34}, 1286 (1975).

\bibitem{Bertsch} G.~Bertsch, S.~J.~Brodsky, A.~S.~Goldhaber and J.~F.~Gunion,
Phys.\ Rev.\ Lett.\  {\bf 47}, 297 (1981).

\bibitem{Jlab_01_107} Jefferson Lab experimental proposal $01-107$, 
K.~Garrow and R. Ent, spokepersons.

\bibitem{JP_8} P. Jain and  J.~P.~Ralston, Phys. Rev. {\bf D49}, 1104 (1993).

\bibitem{oneill} T.~G.~O'Neill {\it et al.},
Phys.\ Lett.\ B {\bf 351}, 87 (1995)

\bibitem{garrow} K.~Garrow {\it et al.},
Phys.\ Rev.\ C {\bf 66}, 044613 (2002).

%
\bibitem{BurMil} M.~Burkardt and G.~A.~Miller,
arXiv:hep-ph/0312190.

\bibitem{EIC} {\it ``Second Electron-Ion Collider Workshop}, Jefferson Lab, 
March 15-17, 2004. \\ $http://www.jlab.org/intralab/calendar/archive04/eic/$
\end{thebibliography}
\end{document}